\documentclass[fullpage,10pt,doublecolumn]{article}

\usepackage{marvosym}

\usepackage{array}
\usepackage{amsmath, amsthm, amssymb}
\usepackage{algpseudocode}
\usepackage{algorithm}
\usepackage{multirow}
\usepackage{bigstrut}
\usepackage{lastpage}
\usepackage{subfig}

  \usepackage[dvips]{graphicx}
  \graphicspath{{./figures/}}
  \DeclareGraphicsExtensions{.pdf}

\usepackage{color}

\newcommand{\Gen}{{G}}
\newcommand{\Acc}{{H}}
\newcommand{\Users}{\mathcal{U}}

\newcommand{\setC}{\mathcal{C}}

\newcommand{\TransProb}{\mathbf{E}}
\newcommand{\TransProbBar}{\bar{\mathbf{E}}}
\newcommand{\UserRegionPref}{\mathcal{P}}
\newcommand{\MobilityPred}{\mathcal{Q}}
\newcommand{\SocialPopularity}{{A}}

\newcommand{\Strategy}{\mathbf{K}}
\newcommand{\Regions}{\mathcal{R}}
\newcommand{\setS}{\mathcal{S}}
\newcommand{\setF}{\mathcal{F}}

\newcommand{\setW}{\mathcal{W}}
\newcommand{\setZ}{\mathcal{Z}}

\newcommand{\PostLambda}{\lambda_{P}}

\begin{document}

\title{Social- and Mobility-Aware Device-to-Device Content Delivery}

\author{
	Zhi Wang,
	Lifeng Sun,
	Miao Zhang,\\
	Haitian Pang,
	Erfang Tian,
	Wenwu Zhu
}

\date{}
\maketitle

\begin{abstract}

	Mobile online social network services have seen a rapid increase, in which the huge amount of user-generated social media contents propagating between users via social connections has significantly challenged the traditional content delivery paradigm: First, replicating all of the contents generated by users to edge servers that well ``fit'' the receivers becomes difficult due to the limited bandwidth and storage capacities. Motivated by device-to-device (D2D) communication that allows users with smart devices to transfer content directly, we propose replicating bandwidth-intensive social contents in a device-to-device manner. Based on large-scale measurement studies on social content propagation and user mobility patterns in edge-network regions, we observe that (1) Device-to-device replication can significantly help users download social contents from nearby neighboring peers; (2) Both social propagation and mobility patterns affect how contents should be replicated; (3) The replication strategies depend on regional characteristics ({\em e.g.}, how users move across regions).
	
	Using these measurement insights, we propose a joint \emph{propagation- and mobility-aware} content replication strategy for edge-network regions, in which social contents are assigned to users in edge-network regions according to a joint consideration of social graph, content propagation and user mobility. We formulate the replication scheduling as an optimization problem and design distributed algorithm only using historical, local and partial information to solve it. Trace-driven experiments further verify the superiority of our proposal: compared with conventional pure movement-based and popularity-based approach, our design can significantly ($2-4$ times) improve the amount of social contents successfully delivered by device-to-device replication.

\end{abstract}

\section{Introduction}

Mobile social network services based on the convergence of wireless networks, smart devices, and online social networks have witnessed a rapid increase in recent years \cite{zhangunderstand}. According to YouTube, over $100$ hours worth of videos have been produced by individuals and shared between themselves, and the traffic of delivering these content items to mobile devices has exceeded $50\%$ \cite{youtube-stat}, significantly challenging the traditional content delivery paradigm, where content is replicated by a hierarchical infrastructure using the same scheme \cite{leighton2003content}. It is usually expansive and inefficient to replicate the massive number of social content items to traditional CDN servers \cite{misloverethinking}.

As the development of the \emph{device-to-device} communication \cite{fodor2012design}, it is promising to offload the bandwidth-intensive social content delivery to users' mobile devices and let them serve each other. Previous studies have demonstrated that such device-to-device content sharing is possible when users are close to each other, and the content to be delivered is delay tolerant \cite{wang2014toss}. In this paper, we use \emph{mobile edge networks} (or edge networks for short) to define the local area where users move across regions and can directly communicate with each other. It is intriguing to investigate content delivery strategies in the context of edge networks, because both users' behaviors and network properties have to be studied.

In traditional device-to-device content sharing, a user usually sends the generated content to a set of users that are close to her in a broadcasting manner, causing the following problems: (1) Due to the broadcasting mechanism, users' devices have to spend expansive power to cache and forward many content items in edge network. 
As the number of user-generated social content items is increasing, such mechanism is inherently in-scalable. (2) Social content---due to the dynamical social propagation---has heterogeneous popularity, while the conventional approaches treat them all the same, resulting in wasted resource to replicate unpopular content items. (3) Due to the dynamic mobility patterns, it is hard to guarantee any quality of experience.

To address these problems, we propose a joint propagation- and mobility-aware replication strategy based on social propagation characteristics and user mobility patterns in the edge-network \emph{regions}, {\em e.g.}, 100x100m$^2$ areas where users can move across and deliver content to other users.

The idea of our proposal is as follows. (1) Instead of letting content flood between users that are merely close to each other, we propose to replicate social content according to the social influence of users and the social propagation of content. (2) We develop a regional social popularity prediction model which captures the popularity of content items based on both regional and social information. (3) We propose to replicate social content items according to not only the regional social popularity, but also user mobility patterns, which capture how users move across and stay in these edge-network regions.

In our proposal, we are facing the following challenges: How to capture the joint propagation and mobility behaviors? How to identify the parameters that affect the performance of mobile social content replication? How to design efficient strategies/algorithms for our proposal to work in real world? Our contributions are a set of answers to tackle these challenges.

$\rhd$ Based on large-scale measurement studies, including $450,000$ content items shared by $240,000$ users on an online social network, and $300,000$ users moving across hundreds of edge-network regions, we reveal the possibility of device-to-device replication for social content, and the design principles to make use of both social propagation and user mobility patterns. We present a number of measurement insights, including how social propagation and mobility patterns affect D2D content replication.

$\rhd$ Based on our measurement studies, we build social propagation and user mobility predictive models, to capture the popularity distribution of content in different regions. Using the predictive models, we then formulate the D2D content replication as an optimization problem, which is inherently centralized. We then design a heuristic algorithm to practically solve it in a distributed manner, which only needs historical, local and partial information. 

$\rhd$ We use both model-driven and trace-driven experiments to verify the effectiveness of our design: compared to traditional approaches, our design can significantly enhance the chance for users to download content from edge-network devices nearby. In particular, our design improves the D2D delivery fraction by $4$ times against a pure movement-based approach, and by $2$ times against a pure popularity-based approach. Based on our model-driven experiments, we also present the limitations of such D2D delivery approach. 

The rest of the paper is organized as follows. We survey related works in Sec.~\ref{sec:relatedwork}. We give the motivation of our design in Sec.~\ref{sec:problem}. Using large-scale measurement studies, we present the principles for our design in Sec.~\ref{sec:measure}. In Sec.~\ref{sec:design}, we present the details of our design based on mobility and propagation predictive models. We evaluate our design in Sec.~\ref{sec:evaluation}. Finally, we conclude the paper in Sec.~\ref{sec:conclusion}.

\section{Related Works} \label{sec:relatedwork}

We survey literature on social propagation, social content distribution, D2D content delivery, as well as user mobility characteristics.

\subsection{Social Propagation in Online Social Networks}

Online social network has greatly changed the content delivery, {\em e.g.}, the distribution of social contents is shifted from a ``central-edge'' manner to an ``edge-edge'' manner. Bakshy \emph{et al.}~\cite{bakshy2011everyone} studied the social influence of people in the online social network, and observed that some users can be very influential in social propagation. Li \emph{et al.}~\cite{livideo2012} studied the content sharing in the online social network, and observed the skewed popularity distribution of contents and the \emph{power-law} activity of users. Comarela et al.~\cite{comarela2012understanding} investigated response time of social contents using collected traces, and confirmed the in nature \emph{delay tolerance} of social media, which motivated our study. In our previous study \cite{zhi-tmm2013}, we also observed the correlation between social connection and propagation, and users' preferences of content.

As online social networks are affecting \emph{dissemination} for all types of online contents, conventional content delivery paradigms need improvement using social information. Pujol \emph{et al.}~\cite{pujol2010little} designed a social partition and replication middleware where users' friends' data can be co-located in datacenter servers. Scellato et al.~\cite{scellato2011track} investigated using social \emph{cascading} information for content delivery over the edge networks. Wang et al.~\cite{zhi-infocom2012,zhi-acmmm2012} investigated the possibility to infer social propagation according to users' social profiles and behaviors, and allocate network resource at edge-cloud servers based on propagation predictions. Wen et al.~\cite{wen2014cloud} further proposed the cloud mobile media concept to utilize cloud-based resource for mobile media content processing and distribution. Wang et al.~\cite{wang2013accelerating} proposed a novel peer-assisted paradigm using social relationship for improved social media distribution.

The limitation of previous studies is that they are focused on social content delivery using server-based hierarchical infrastructure. Our study will investigate how bandwidth-intensive social contents can be distributed by D2D resources, by jointly infer social propagation and user mobility.

\subsection{Mobility Characteristics}

Understanding mobility of users is a key to design effective delivery strategies for mobile social contents. Kim et al.~\cite{kim2006extracting} proposed to use traces of users' associations with Wi-Fi access points to investigate how users move among popular locations. Based on user mobility models, Yoon et al.~\cite{yoon2006building} found that it is possible to generate movement patterns which are statistically similar to the real movement. Karamshuk et al.~\cite{karamshuk2011human} surveyed the usage of spatial, temporal and social properties to capture the mobility behaviors. 

Rhee et al.~\cite{rhee2011levy} then pointed out that human movements are not random walks, and the patterns of human walks and \emph{Levy} walks contain some statistical similarity; in particular, features including heavy-tail flight and the super diffusive nature of mobility are observed. In \cite{karagiannis2010power}, power-law and exponential decay of inter-contact times between mobile devices are observed. Zhuang et al.~\cite{zhuang2012inferring} studied the mobility and encountering patterns for users in regions during particular events, {\em e.g.}, conferences.

When jointly studying mobility and social network of users, Cho et al.~\cite{cho2011friendship} observed that though human movement and mobility patterns have a high degree of variation, they exhibit structural patterns due to geographic and social constraints. In particular, short-ranged travel is periodic both spatially and temporally, and not affected by the social network structure. Recently, such mobility studies have improved the edge network and content delivery design, Wang et al.~\cite{wang2010efficient} studies the mobility characteristics of people to guide wireless network deployment. Wang et al.~\cite{wang2011human} found the similarity between individuals' mobility patterns and their social proximities.

These studies have focused on human mobility characteristics from a general way, {\em i.e.}, how people move in their daily lives. Our study will particularly focus on how users move in edge-network regions, {\em e.g.}, regions associated with Wi-Fi access points, where users can serve as peers for D2D content delivery.

\subsection{Message Forwarding in Delay-tolerant Networks}

The delay-tolerant network architecture and application interface was proposed to structure around optionally-reliable asynchronous message forwarding \cite{fall2003delay}, with limited expectations of end-to-end connectivity and node resources. Since then, many efforts have been devoted to efficient message routing and forwarding in this paradigm. Jain et al.~investigated the routing problem in such delay-tolerant networks, and provided that algorithms using the least knowledge tend to perform poorly, while with limited additional knowledge, far less than complete global knowledge, efficient algorithms can be constructed for routing in such environments \cite{jain2004routing}. Daly et al.~studied the small world dynamics for characterizing information propagation in wireless networks, and also confirmed that using local information is promising for message routing in DTNs \cite{daly2007social}. Helgason et al.~\cite{helgason2010mobile} developed an opportunistic framework, in which content is divided into different topics for lookup and forward. Haillot et al.~\cite{haillot2010protocol} proposed a content-based communication scheme, where users can subscribe to content categories according to their preferences, and content is disseminated based on the subscription. Hui et al.~\cite{hui2011bubble} studied the patterns of contact in pocket switched networks, and exploited two social and structural metrics: centrality and community, using real human mobility traces. Karamshuk et al.~\cite{karamshuk2011human} surveyed the usage of spatial, temporal and social properties to capture the mobility behaviors, and utilize mobility models for opportunistic networks.

In these studies, the concept of social network is used as real-world contact graph, i.e., the contact of people forms a social network. However, today's online social networks and the user contact graph can be highly independent on each other, as users are allowed to interact with any users that are not necessarily around, thanks to the wireless network and mobile devices. Our study is to investigate how social propagation and mobility patterns can be jointly utilized to improve content replication in edge networks, and the limitations of such D2D replication for social media content.

\subsection{Mobile Social Content Delivery}

As early as Mobile Ad hoc Network (MANET) was proposed, researchers started to envision using device-to-device communications for content delivery. Pelusi et al.~\cite{pelusi2006opportunistic} studied opportunistic networking and discussed possible scenarios to use it. In the context of using smart devices with limited energy for content delivery, Li et al.~\cite{li2010energy} formulated the optimization problem of opportunistic forwarding, with the constraint of energy consumed by the message delivery for both two-hop and epidemic forwarding. In Han et al.'s study \cite{han2012mobile}, they investigated the selection of target-set of users for content deployment to minimize the mobile data traffic over cellular networks.

Recently, Bao et al.~\cite{bao2013dataspotting} have explored the possibility of serving user requests from other mobile devices located geographically close to the user, where the cellular operator learns users' content request, and guides smart devices to other devices that have the requested content. Wang et al.~\cite{wang2014toss} have investigated the problem of offloading social media delivery based on a jointly online social network and mobile social network framework.

In previous studies, it is assumed that users directly help each other in their \emph{trajectories}, which significantly limits the ability of D2D delivery possibility. In our study, we propose to replicate social contents for regional social propagation, {\em i.e.}, users are supposed to serve edge-network regions where they will move to and stay, based on joint social propagation and user mobility characteristics, so that edge-network regions will be covered by peering users, according to their regional social popularities.

\section{Motivation and Assumption} \label{sec:problem}


\subsection{Motivation}

In an online social network, users share contents with their friends through the social connections. The propagation of contents is determined by both social graph and user behaviors. As such social graph and user behaviors are inside the online social network, they can be independent to users' mobility patterns in the physical world. For example, a user can intensively interact with her friends online without having to be at the same location with them, thanks to the online social network.

\begin{figure}[!t]
	\centering
		\includegraphics[width=\linewidth]{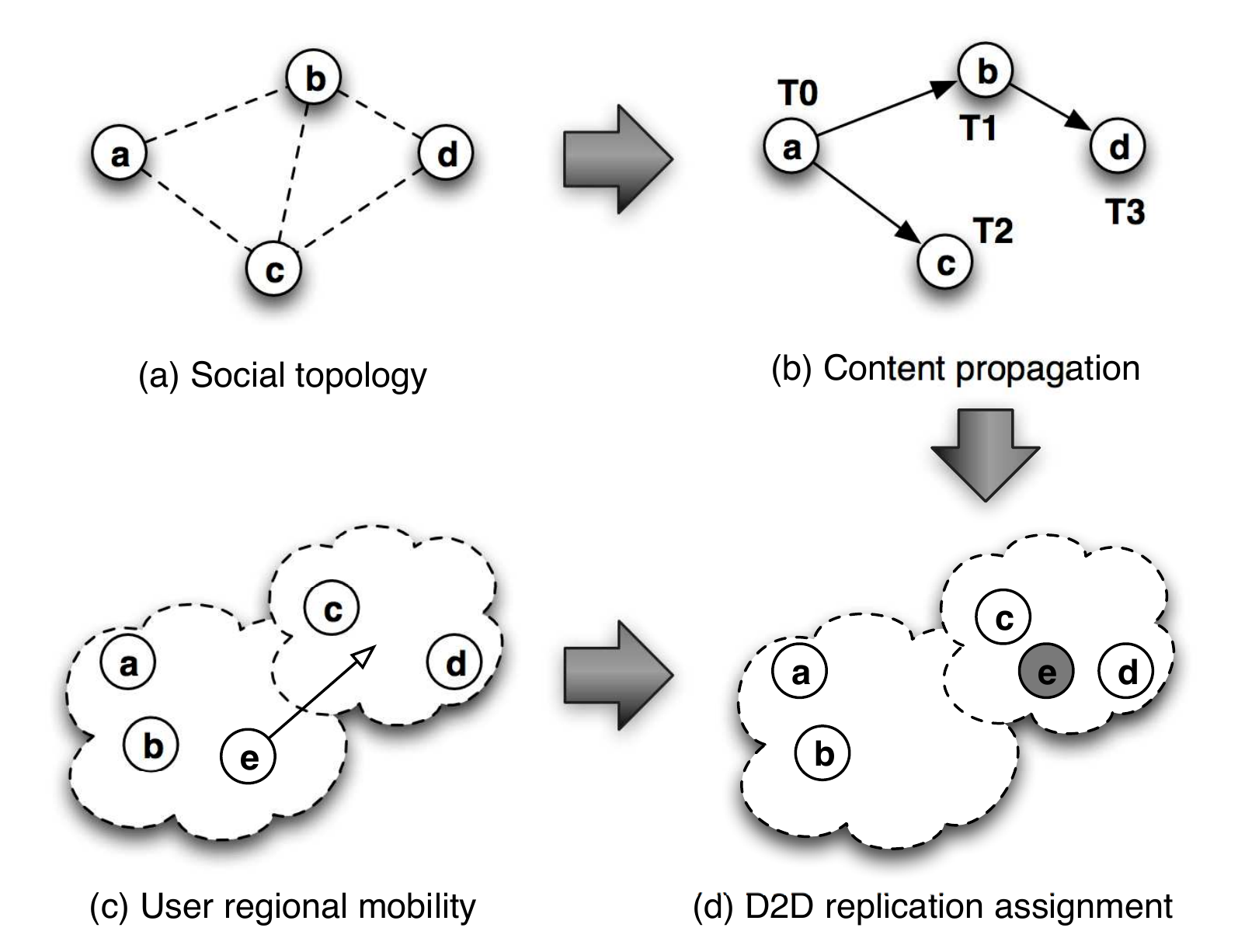}
		\caption{D2D replication affected by social topology, content propagation and user regional mobility.}
	\label{fig:user-topologies}
\end{figure}

As illustrated in Fig.~\ref{fig:user-topologies}, based on the social graph and propagation patterns, we estimate how contents will be received by users; based on the regional mobility, we will then predict which regions users will be moving to and how long they will stay. Jointly, we decide which users will replicate which social contents on the move. In this example, as user $e$ -- while not a friend to any other user -- is moving to the region where user $c$ and $d$ are. Thus, $e$ will be selected to replicate the content generated by user $a$, and both user $c$ and $d$ will receive the content shared by user $a$ in the social propagation at times $T2$ and $T3$, respectively.

In our study, we propose to jointly make use of the social graph, user behaviors and user mobility patterns for D2D replication, to serve edge-network regions. To this end, we first study propagation of mobile social contents, users' mobility patterns, and characteristics of edge-network regions. Based on the measurement insights, we then propose a propagation- and mobility-aware D2D replication for edge-network regions.

\subsection{Assumption}

In our proposal, we decouple the content replication for the social propagation the online social network, from users' mobility in the physical world, i.e., a user may cache content and deliver the content to other users who are not socially connected to her. Be noted that though our replication is decoupled from the mobility, the social propagation and mobility patterns do not have to be independent, e.g., users may share more content at some locations than others. In our design, we present an algorithm to adaptively select content for users to cache. We will also verify the impact of the correlation between the social propagation and user mobility in our experiments.

\section{Measurement Studies} \label{sec:measure}

We carry out large-scale measurement studies on social propagation, user mobility and edge-network regions.

\begin{figure*}[!th]
	\begin{minipage}[t]{.48\linewidth}
		\centering
			\includegraphics[width=.85\linewidth]{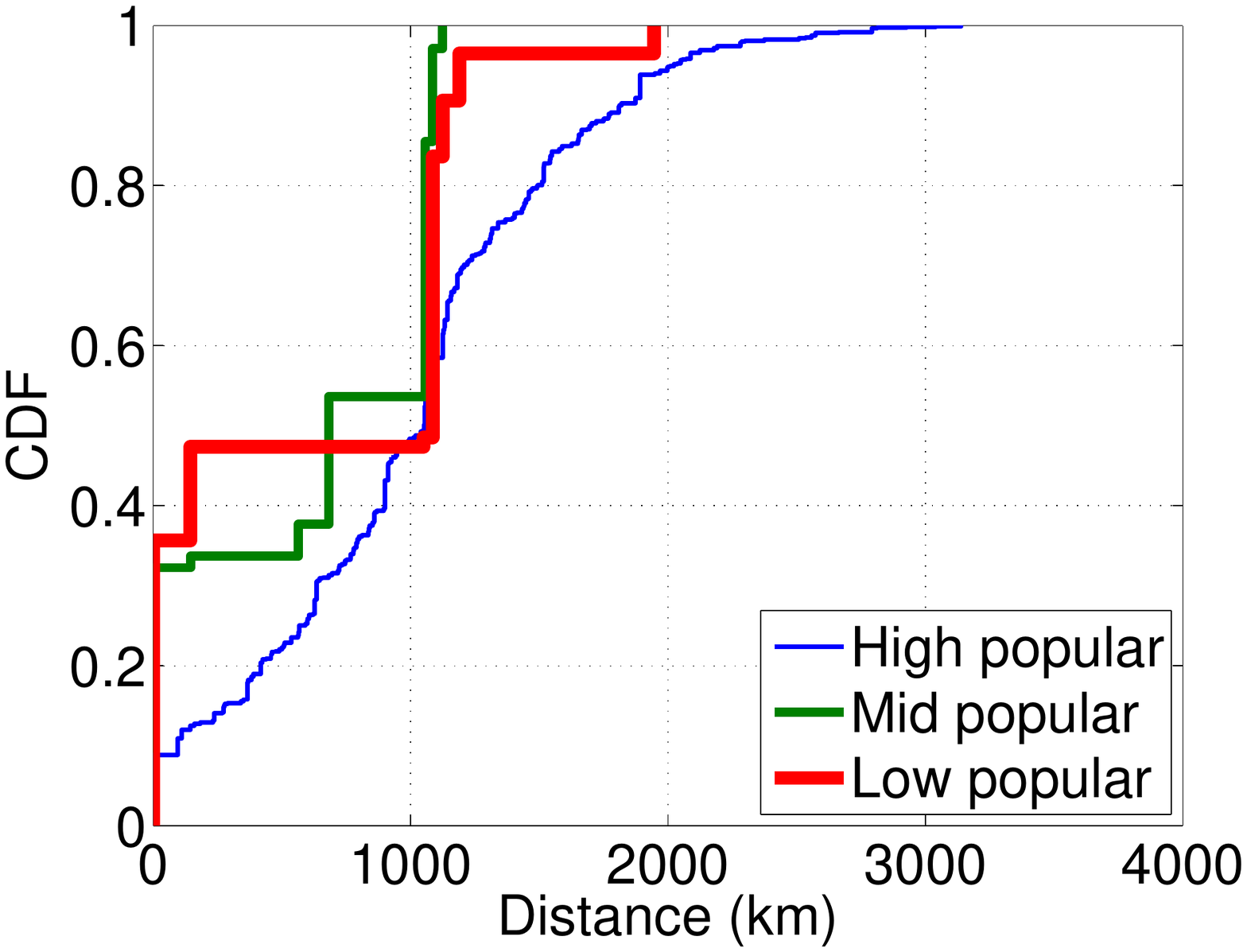}
		\caption{CDF of distance between users that are sharing in the same propagation of social content items.}
		\label{fig:distance-cdf}
	\end{minipage}
	\hfill
	\begin{minipage}[t]{.48\linewidth}
		\centering
			\includegraphics[width=.85\linewidth]{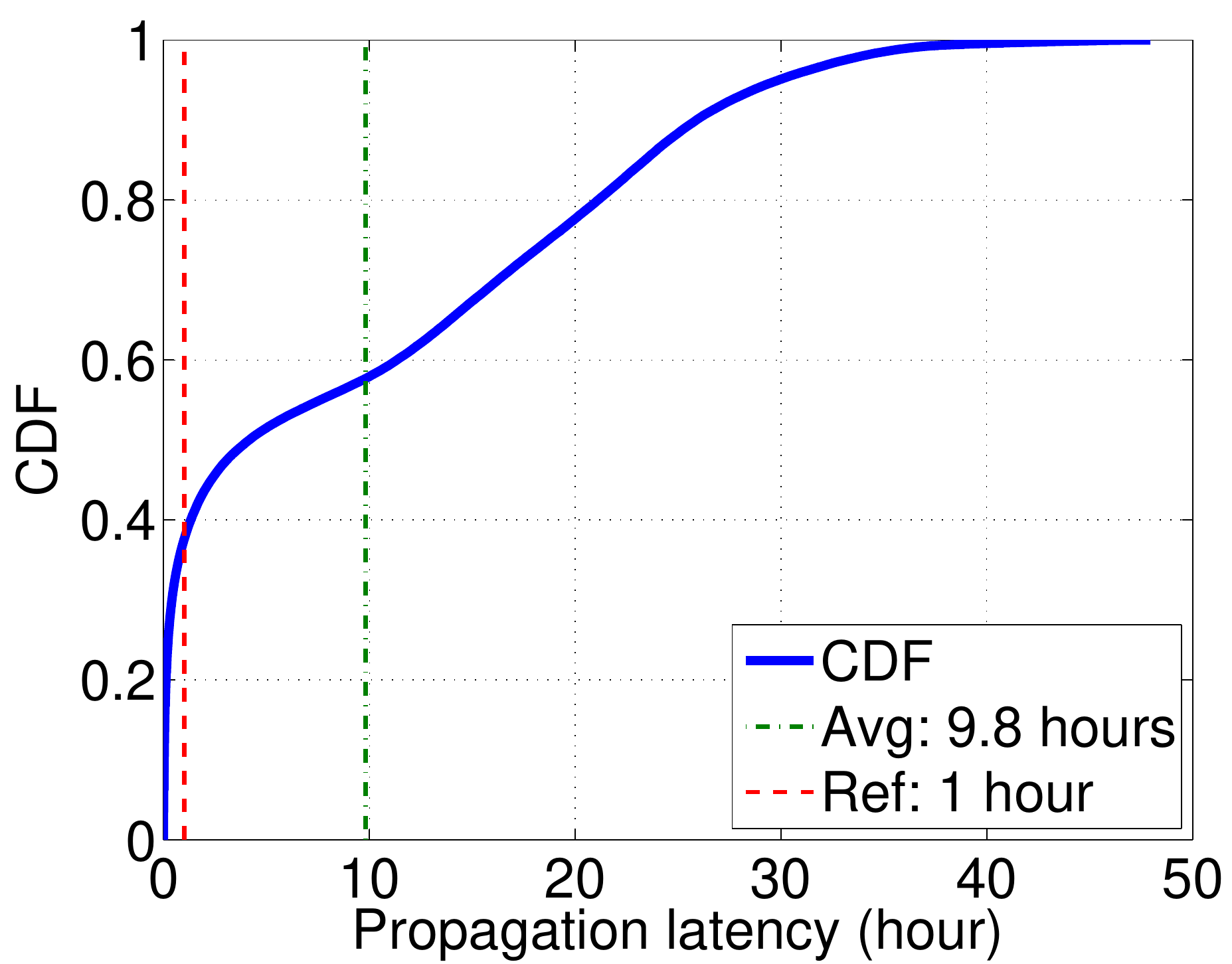}
		\caption{CDF of propagation latency.}
		\label{fig:delay-tolerance}
	\end{minipage}
\end{figure*}

\subsection{Measurement Setup}

We use a data-driven approach for the measurement study, using the following valuable datasets provided by our partners. 

\subsubsection{Traces of Mobile Social Content Propagation}

We have obtained content upload and request traces from partners of Tencent Weishi, a mobile social network service. In Weishi, short videos are generated by individuals and shared with their friends. Our dataset records $240,000$ users sharing over $450,000$ videos in 2014, with the following information: (1) Content generation, which records when a video is generated and shared by a particular user; (2) Content download, which records which videos are downloaded by which users; (3) Social graph, which records how users are socially connected to each other; (4) Sharing information, which records when content is shared by users, including the ID, name, IP address of the users, time stamp when a content is shared, and IDs of the parent and root users if it is a reshare.

\subsubsection{Traces of Edge-network Region and User Mobility}

To investigate how users move across edge-network regions, we use traces provided by NextWiFi, a local Wi-Fi provider. NextWiFi has sampled over $300,000$ users associating to hundreds of Wi-Fi access points in a shopping mall in 2014. The information NextWiFi recorded includes (1) The timestamp when users connect to Wi-Fi access points, and the duration of the AP \emph{association}; (2) The BSSID, Service Set Identification (SSID) and locations of these access points.

Based on these datasets, we first study how social content items are shared by users moving across edge-network regions, and the measurement studies will reveal the possibility of D2D social content replication; Then, we study users' mobility patterns, which reveal the principles to design the replication strategies.

\begin{figure*}[t]
	\begin{minipage}[t]{.48\linewidth}
		\centering
			\includegraphics[width=.85\linewidth]{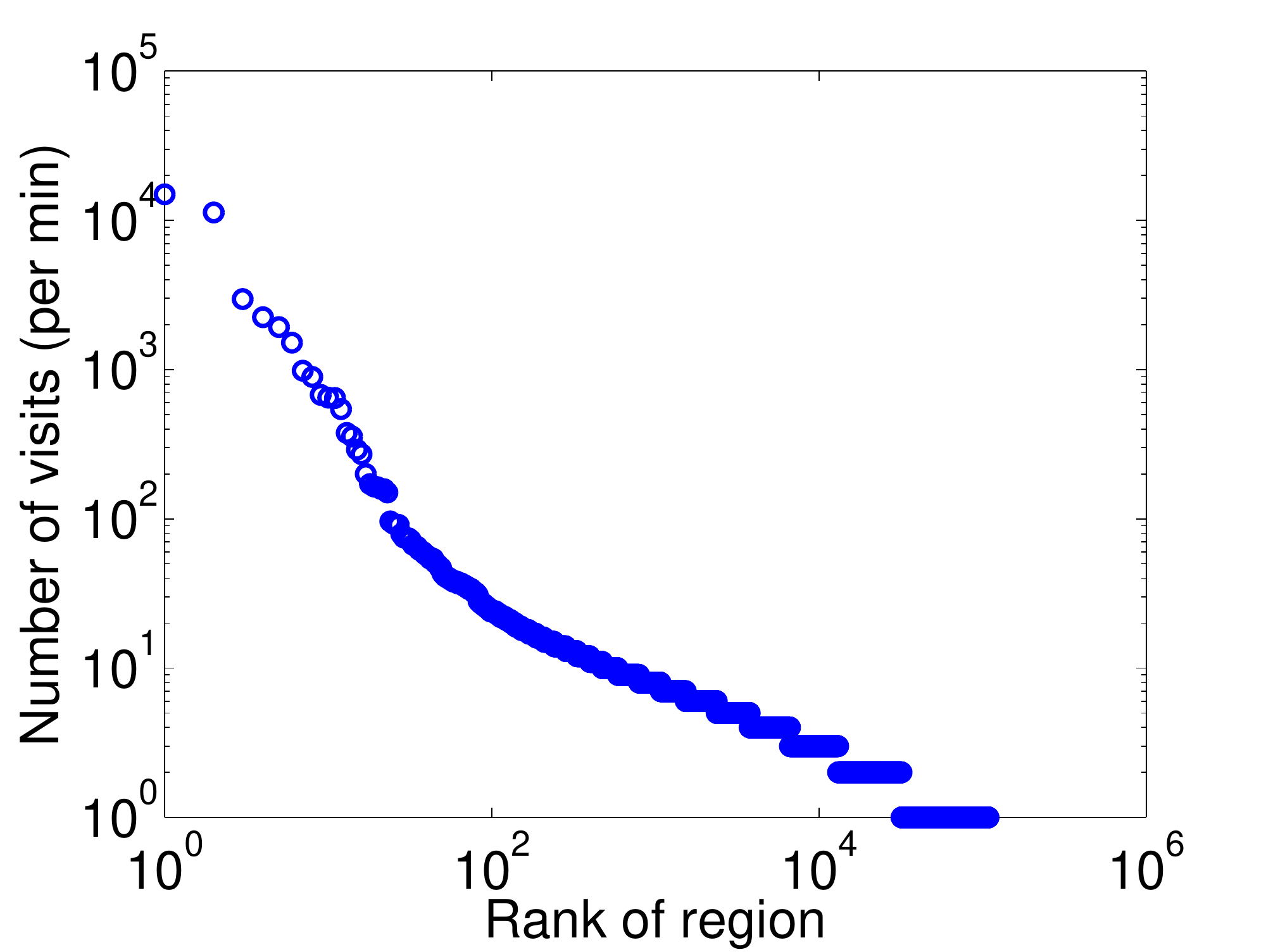}
		\caption{Number of users' visits to edge-network regions.}
		\label{fig:region-popularity}
	\end{minipage}
	\hfill
	\begin{minipage}[t]{.48\linewidth}
		\centering
			\includegraphics[width=.85\linewidth]{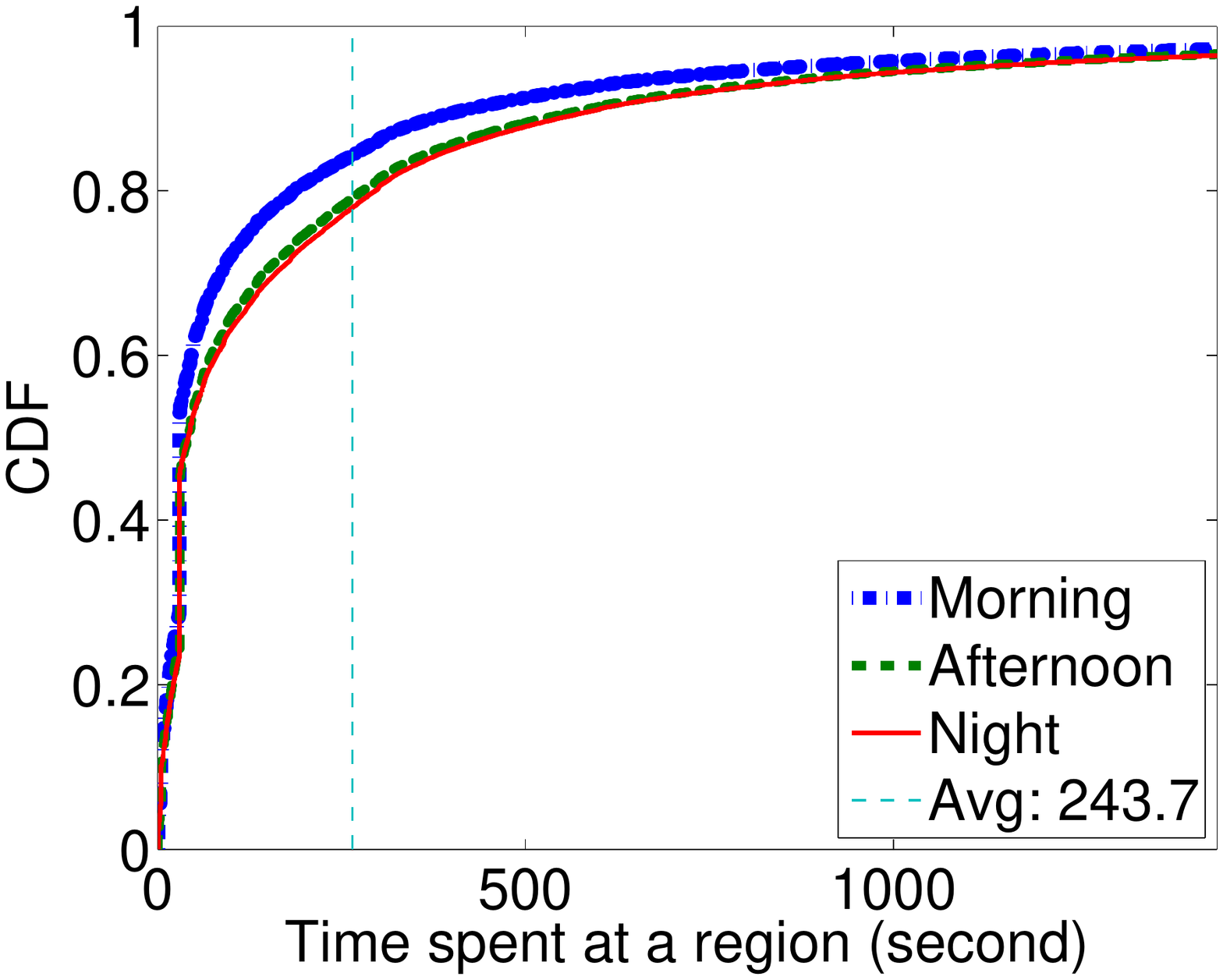}
		\caption{CDF of duration of user association in edge-network regions.}
		\label{fig:associate-duration}
	\end{minipage}
\end{figure*}

\subsection{Propagation of Social Content}

\subsubsection{Propagation in Local Regions}

We first investigate the possibility of D2D content delivery for social content based on the propagation traces. In Fig.~\ref{fig:distance-cdf}, we plot the CDF of distances between users who join the social propagation of the same content in the online social network. Different curves are for content items with different popularities, {\em i.e.}, the number of users who receive the content. We observe that different from popular content that is shared by users randomly located in different places, unpopular content is more likely to be shared in \emph{local} regions, where users are more close to each other. For example around $40\%$ of the distances between users sharing the unpopular content (low popular) are close to $0$km, indicating that users may be in the same edge-network regions.

This observation indicates that in propagations of such unpopular content, which are the majority of social content \cite{cha2007tube}, users are likely to be able to receive content by a D2D scheme. Notice that these distances are between users sharing the same social content; in our design, we also allow users who are in different content propagations to deliver content for each other, which can further reduce the distance between users for D2D delivery.

\subsubsection{Delay Tolerance of Receiving Social Content}

D2D content delivery depends on users who are moving across edge-network regions, {\em e.g.}, a user can carry a social content from one region and then serve it in another region when the user moves there. As a result, D2D content delivery itself may not be able to guarantee a latency to successfully pass a content due to dynamics of user mobility, since the content has to wait for users to encounter each other to be delivered.

Thus, we have to investigate whether it is feasible for today's social content delivery, from the perspective of latency tolerance of users. We study a \emph{propagation latency}, which defines the time elapse between the time when a content is shared by a person and the time when the content is reshared or viewed by another user. D2D delivery is supposed to be carried out within this propagation latency.

As illustrated in Fig.~\ref{fig:delay-tolerance}, the curve represents the CDF of propagation latencies of social content shared in our traces. In particular, the average propagation latency is around $9.8$ hours, and over $62\%$ ({\em resp.} $69\%$) of the propagation latencies are larger than $1$ hour ({\em resp.} $30$ minutes). These observations indicate that in social propagation, social content tends to be delay tolerant, allowing us to design D2D replication for content delivery.

\subsection{Characteristics of Edge-network Regions}

Next, we study the characteristics of edge-network regions, where users are supposed to replicate social content and serve others in our design.

\subsubsection{Popularity of Edge-Network Regions} \label{sec:region-popularity}

We first study the popularity of edge-network regions, in terms of the number of users in different regions. Several previous works assumed users' movement as random walks \cite{rhee2011levy}, which is however not true for users' daily mobility patterns. In our study, we investigate the popularity of different edge regions, which captures how many users are currently in these regions, who can be either requesting users or D2D peers.

As illustrated in Fig.~\ref{fig:region-popularity}, each sample is the number of visits per minute versus the rank of regions (in descending order of the number of visits). We observe that the popularity roughly follows a \emph{zipf}-like distribution, indicating that there are a few regions that can attract much more users than others. In our design, this observation is taken into account, so that a region with a higher ``load'' ({\em i.e.}, the number of requests that can potentially be issued from a region) will be assigned more D2D peers to replicate the social content.

\subsubsection{Association Duration in Different Regions} \label{sec:ass-duration}

After leaving an edge-network region, a user will no longer be able to serve users in that region. Thus, it is also important to investigate how long users can stay in a region. From the Wi-Fi traces, we have sampled $5$ million Wi-Fi association records spanning $10$ days, to study the association duration of users in the regions they have visited. In Fig.~\ref{fig:associate-duration}, we plot the CDF of time users stay in regions they visited, at different hours of a day. The average duration for users to stay in a region is around $4$ minutes, and we also observe that users tend to spend less time in the same region in the morning than in other hours. These observations indicate that there are still many users that only instantly pass edge-network regions, and are probably not able to serve as stable replication peers. In our design, we take how long users stay in different regions into account.

\subsection{Regional Mobility Patterns} \label{sec:user-pref-region}

Finally, we study users' mobility patterns in edge-network regions, to guide our D2D replication design.

\subsubsection{Revisit to the Same Regions} 

\begin{figure*}[t]
	\begin{minipage}[t]{.48\linewidth}
		\centering
			\includegraphics[width=.85\linewidth]{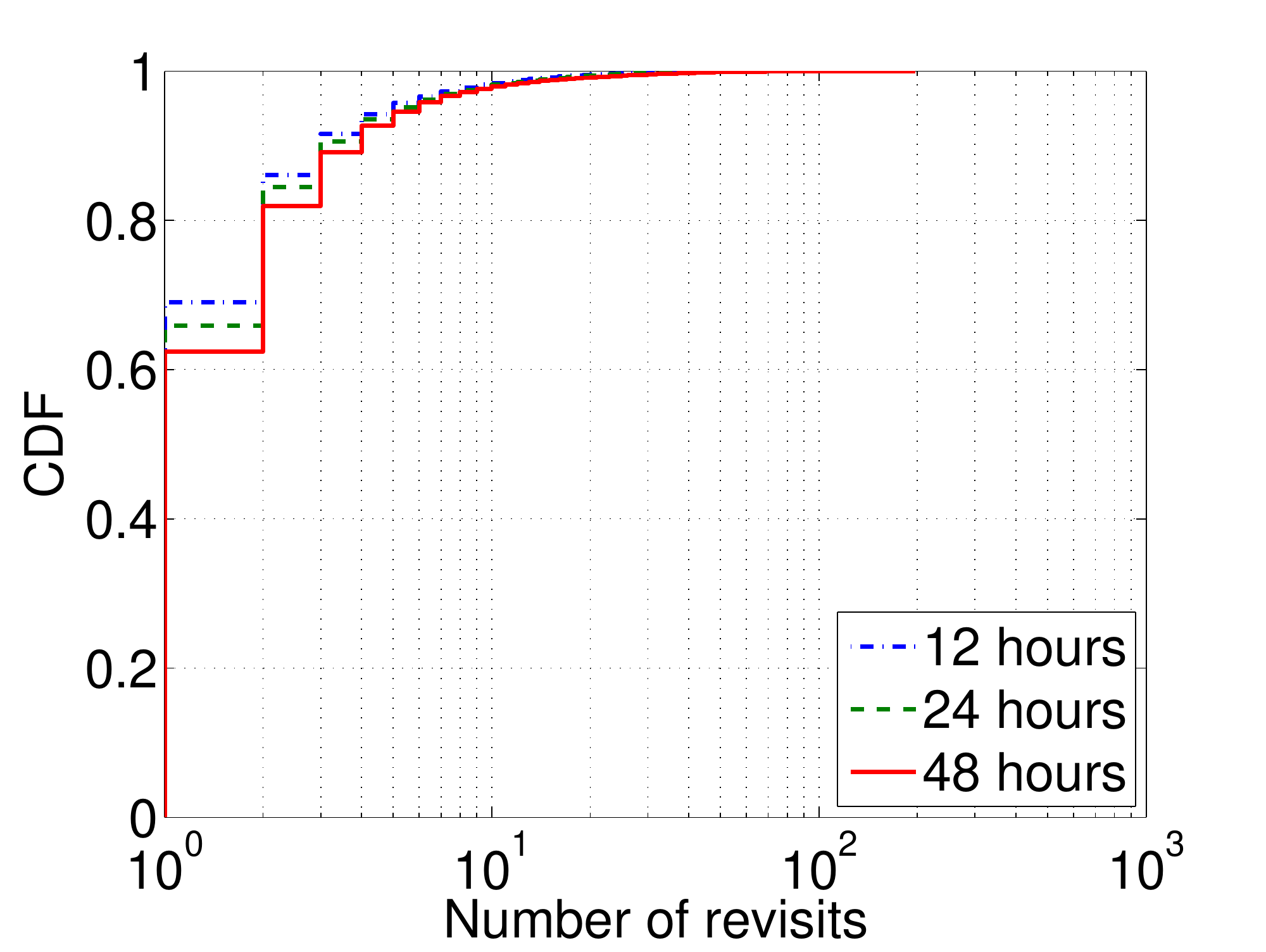}
		\caption{CDF of the number of users' visits to the same region.}
		\label{fig:revisit-cdf}
	\end{minipage}
	\hfill
	\begin{minipage}[t]{.48\linewidth}
		\centering
			\includegraphics[width=.85\linewidth]{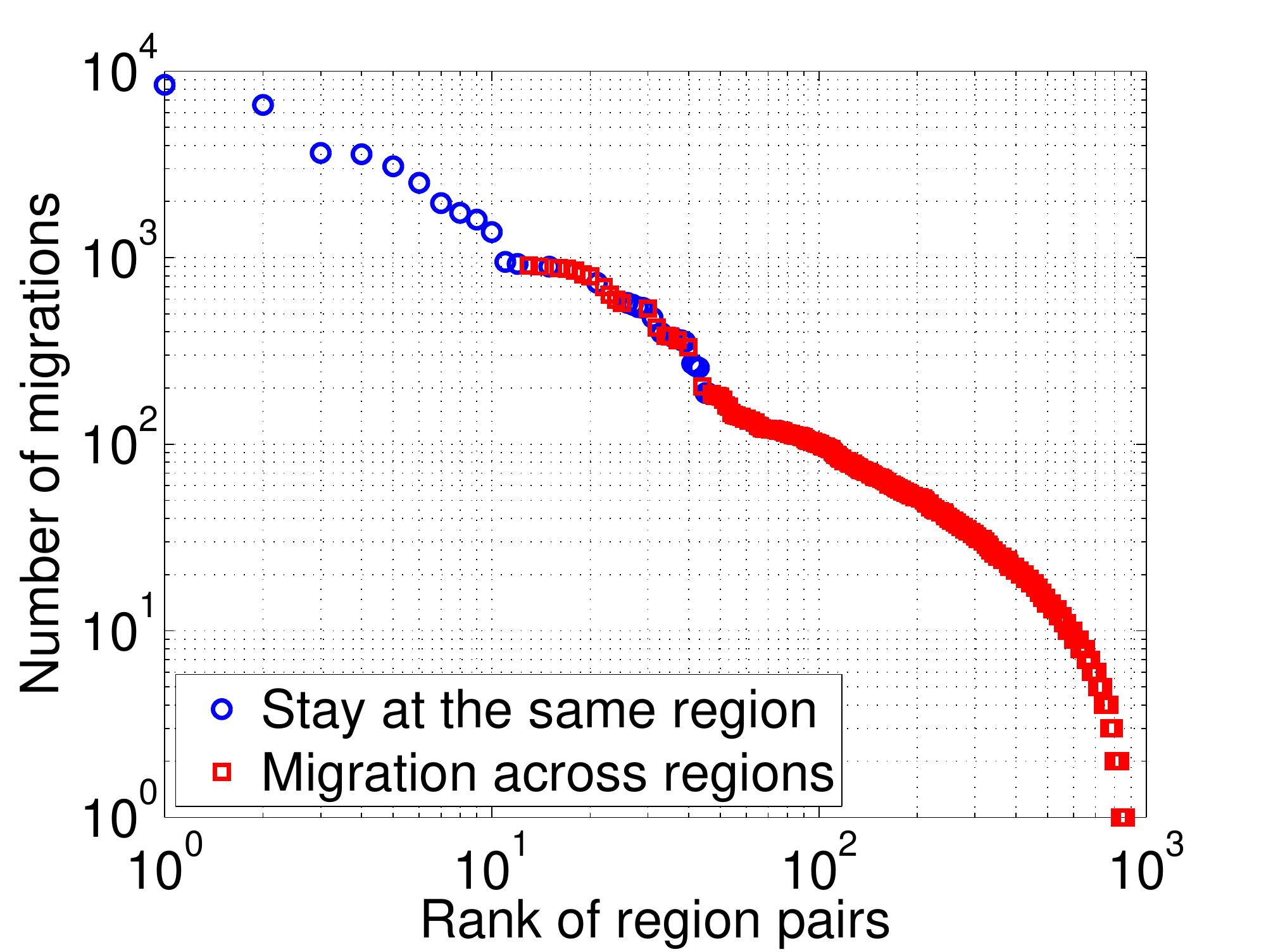}
		\caption{Number of user migrations across regions.}
		\label{fig:region-migration}
	\end{minipage}
\end{figure*}

We investigate the possibility for users to revisit the same region for many times, as this helps us to design social content replication strategies according to users' periodical appearance patterns. We study how users revisit the same region using the Wi-Fi traces we collected. In Fig.~\ref{fig:revisit-cdf}, the curves are CDFs of number of users' revisits to the same region within a time span of $12$ hours, $24$ hours and $48$ hours, respectively. We observe that over $30\%$ of users' visits happen to the same region over $2$ times. This observation indicates that users are actually not randomly walking across edge-network regions; instead, they have inherent preferences of different regions. As a result, replicating social content should take this region preference into consideration.

\subsubsection{Migration Patterns between Regions} \label{sec:migration}

Furthermore, we investigate how users migrate between different regions. For each region pair $(r1,r2)$, we calculate a migration number of users who connect to Wi-Fi hotspots in $r1$ and $r2$, respectively, in two consecutive associations. Without loss of generality, a user associates with APs in the same region if $r1=r2$. We then rank the region pairs in descending order of the migration number. In Fig.~\ref{fig:region-migration}, we plot the migration number versus the rank of sorted region pairs. Notice that users can associate with the same Wi-Fi access point for many times, thus there are ``migrations'' staying in the same region. We observe that this number roughly follows a power-law distribution, indicating that there are a few pairs of regions that users tend to move across each other.

This observation indicates that edge-network regions are actually ({\em e.g.}, geographically) correlated with each other. Based on this observation, we propose to predict which region a user may move to in the future according to the migration patterns in our design.

\section{Detailed Design: Propagation- and Mobility-Aware D2D Regional Replication} \label{sec:design}

\begin{figure}[t]
	\centering
		\includegraphics[width=0.7\linewidth]{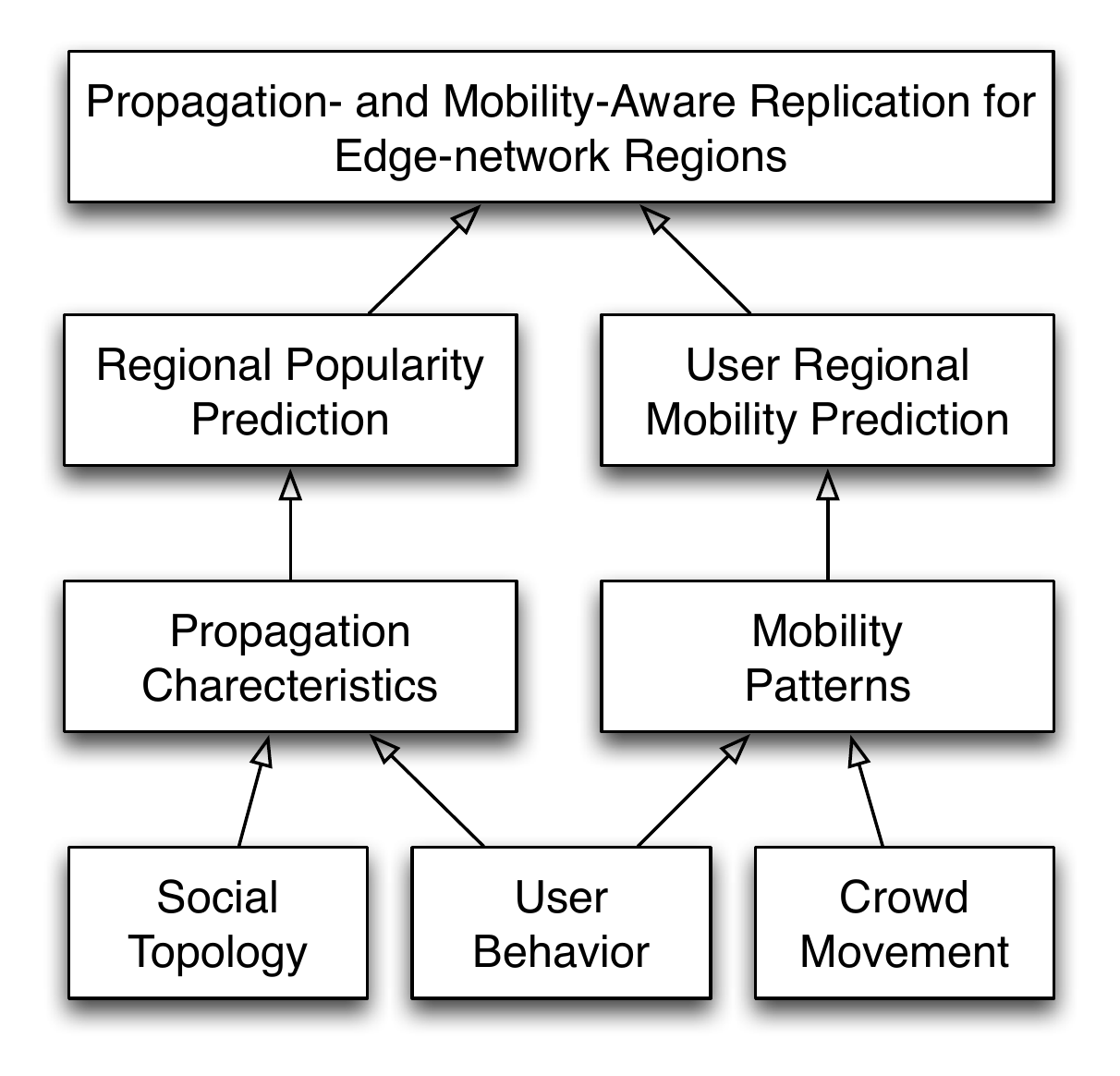}
	\caption{Framework of propagation- and mobility-aware D2D replication for social content.}
	\label{fig:framework}
\end{figure}

Motivated by the measurement insights, we design a joint propagation- and mobility-aware replication strategy for D2D social content delivery. The framework of our design is illustrated in Fig.~\ref{fig:framework}. We design models to capture the propagation patterns of social content, and the regional popularity and user mobility across regions. Based on these models, we predict if a content will be highly requested due to social propagation in a particular region, and we predict where a user may visit in the future. Using these predictions, we assign users to replicate social content for D2D delivery.

Before we present the details, we give some important notations in Table \ref{tab:notations}.
\begin{table}[!t]
	\caption{Important notations}
	\label{tab:notations}
	\renewcommand{\arraystretch}{1.3}
	\centering
	\begin{tabular}{|p{0.12\linewidth}||p{0.78\linewidth}|}
		\hline
			Symbol & Definition \\
		\hline
			\emph{$u,v$} 			& User index\\
			\emph{$r,s$} 			& Region index\\
			\emph{$\SocialPopularity_{c,r}^{(T)}$} 			& Regional social popularity of content $c$ in region $r$\\
			\emph{$K_{u,c}^{(T)}$} 		& Strategy variable indicating whether user $u$ will carry $c$ for D2D replication\\
			\emph{$\MobilityPred_{u,r}^{(T)}$} 		& Probability for user $u$ to visit region $r$ \\
			\emph{$I_{u,v}^{(T)}$}	& A social influence index from user $u$ to $v$\\
			\emph{$\Acc_{u,v}^{(t)}$} & The number of content items that are shared by user $u$ and received (accepted) by user $v$ in time slot $t$\\
			\emph{$\Gen_u^{(t)}$} 	& The number of contents posted by user $u$ in time slot $t$\\
			\emph{$\UserRegionPref_u^{(T)}$} 	& {A user's preference vector of different regions.}\\
			\emph{$\TransProb_{r,s}^{(T)}$} 	& {Number of user migrations from region $r$ to $s$.}\\
		\hline
	\end{tabular}
\end{table}

\subsection{Propagation-based Regional Popularity}

We aim to characterize the popularity of social content. In particular, we seek to answer the following question: How many users in total are expected to download a social content in a particular edge-network region in the next time slot? Notice that we propose to schedule D2D replication strategies in a discrete time slot-based manner, and there are plenty of existing studies to convert such strategies to real implementation \cite{zhang2007understanding}. The length of a time slot depends on the observed average propagation latency and the duration users staying in the regions.

%
%

\subsubsection{Social Influence of Users}

We propose to build an \emph{influence} estimation between users for regional propagation prediction. Let $I_{u,v}^{(T)}$ represent a social influence index from user $u$ to $v$, which captures the ability for user $u$ to attract user $v$ to receive content shared by user $u$ in time slot $T$. $I_{u,v}^{(T)}$ is affected by many factors \cite{cui2011should}, including the content itself, the preference of users, and the context under which the content is shared.

In our design, we use a data-driven approach to estimate the influence index, by considering the historical statistics as follows: 
\begin{equation} 
	I_{u,v}^{(T)} = \frac{\sum_{t=T-W}^{T-1} \Acc_{u,v}^{(t)}}{\sum_{t=T-W}^{T-1} \Gen_u^{(t)}}, 
\end{equation}
where $\Acc_{u,v}^{(t)}$ is the number of social content items that are shared by user $u$ and received (accepted) by user $v$ in time slot $t$, and $\Gen_u^{(t)}$ is the number of all content generated and reshared by user $u$ in time slot $t$. $W$ is a time window that we are referring to for statistics, and is chosen to be large enough to contain the content shared by user $u$. The rationale is that we use users' social influence recorded in previous time slots to estimate how much they attract others in the future. A larger $I_{u,v}^{(T)}$ indicates that user $v$ is more likely to accept content shared by user $u$. If user $u$ has a large influence index to many friends, the popularity affected by social propagation then cannot be ignored.

\subsubsection{User Regional Preference}

According to our measurement studies in Sec.~\ref{sec:user-pref-region}, users have different preferences of different edge-network regions, reflecting possibilities where they stay. We investigate users' inherent preference of different regions. Suppose a user has a preference $\UserRegionPref_{u,r}^{(T)}$ of a region $r$. We estimate $\UserRegionPref_{u,r}^{(T)}$ using the records of how long users are staying in different regions in a previous time window $W'$, as follows:
\begin{equation}
	\UserRegionPref_{u,r}^{(T)} = \frac{d_{u,r}^{(T-W',T-1)}}{\sum_{r' \in \Regions_u^{(T)}}d_{u,r'}^{(T-W',T-1)}},
\end{equation}
where $d_{u,r}^{(T-W',T-1)}$ is the duration user $u$ has stayed in region $r$ in the previous time window $[T-W', T-1]$, $\Regions_u^{(T)}$ is the set of regions user $u$ has visited before, and $W'$ is the time window to study users' regional preference. In our design, $W'$ is chosen on a weekly basis, {\em i.e.}, the previous week is referred to for estimation.

\subsubsection{Regional Social Popularity}

Based on both the user influence and users' region preference, we study the regional social popularity.

$\rhd$ \emph{Inherent content popularity}. In our design, the inherent content popularity is determined by the content itself. Some content items tend to be more interesting and can attract more attentions than others. We use $p_{c,r}^{(T)}$ to denote the inherent popularity of content $c$ in region $r$ in time slot $T$, which can be inferred using traditional popularity prediction approaches \cite{szabo2010predicting}.

$\rhd$ \emph{Influential popularity}. After a content is distributed over the social connections, its popularity is highly affected by the social networks \cite{zhi-acmmm2012}. We thus incorporate the social influence of users into our popularity inference.

Based on the inherent popularity and influential popularity, we design a \emph{social popularity index} $\SocialPopularity_{c,r}^{(T)}$, which reflects the popularity of content $c$ in region $r$ in time slot $T$, as follows:
\begin{equation}
	\SocialPopularity_{c,r}^{(T)} = p_{c,r}^{(T)} + \alpha \sum_{u \in \setS(c)} \sum_{v \in \setF_u} I_{u,v}^{(T)} \UserRegionPref_{v,r}^{(T)},
\end{equation}
where $\setS(c)$ is the set of users who have shared content $c$, and $\setF_u$ is the set of users who are socially connected to user $u$, which may become the resharers next. 
According to our previous definitions, $\sum_{u \in \setS(c)} \sum_{v \in \setF_u} I_{u,v}^{(T)} \UserRegionPref_{v,r}^{(T)}$ reflects the popularity of social influence in region $r$.
$\alpha \in [0,1]$ is a running parameter to determine how much the social influential popularity can contribute to the content requests. For example, some content items are only shared between friends and the social influence thus dominates the popularity; while some other content items are published by the central content provider and their popularity is affected by the content itself \cite{cui2011should}.

$\alpha$ can be learnt by historical data: $\alpha$ is assigned a larger value for content whose popularity is more influenced by the social propagation, {\em i.e.}, the correlation between content popularity and social influence is strong. In our experiments, we use the fraction of the influenced users ({\em i.e.}, users whose friends are already resharers of a content) over all viewers to represent such inference level of a particular content.

A content with a large $\SocialPopularity_{c,r}^{(T)}$ is likely to attract more requests from region $r$. The rationale is that a content with a large social popularity index is either very popular in a region in the previous time window, or has been shared by many influential people whose resharers are located in the region.

\subsection{Regional Mobility Prediction} \label{sec:mobility-prediction}

\subsubsection{Regional Migration Model}

We propose to let users cache content and serve as edge peers in local regions. To this end, we have to understand how users move across different regions. Based on our previous measurement studies, we observe that the edge regions not only have different popularities, but also demonstrate different correlation levels between each other -- more users tend to migrate between some pairs of regions than others.

Let $\TransProb_{r,s}^{(T)}$ denote the number of migrations of users who have moved from region $r$ directly to region $s$ in the previous time slot $T-1$ ({\em i.e.}, only the recent previous time slot is referred to). We normalize $\TransProb_{r,s}^{(T)}$ to $\bar{\TransProb}_{r,s}^{(T)}$, as below: $ \TransProbBar_{r,s}^{(T)} = \frac{\TransProb_{r,s}^{(T)}}{\sum_{r'|r' \in \Regions} \TransProb_{r's}^{(T)}}, $ where $\Regions$ is the set of all edge-network regions. In our later formulation of the replication strategy, we use this \emph{migration index} $\TransProbBar_{r,s}^{(T)}$ as an optimization coefficient.

\subsubsection{User Regional Mobility Prediction}

Based on the user regional migration index and the regional preference of users, we then study a user's \emph{mobility index} $\MobilityPred_{u,r}^{(T)}$ to capture the possibility for user $u$ to visit region $r$ in time slot $T$ from its current region. The calculation of user mobility index is defined as below:
\begin{equation}
	\MobilityPred_{u,r}^{(T)} = 
	\frac{\TransProbBar_{R_u^{(T-1)}, r}^{(T)} \UserRegionPref_{u,r}^{(T)}}
		 {\sum_{s \in \Regions} \TransProbBar_{R_u^{(T-1)},s}^{(T)} \UserRegionPref_{u,s}^{(T)}},
\end{equation}
where $R_u^{(T-1)}$ is the region where user $u$ is in the previous time slot ({\em i.e.}, $T-1$), and $\UserRegionPref_{u,r}^{(T)}$ is a user's region preference defined before.
The rationale of the user mobility index is that it jointly uses the user mobility statistics, which capture the correlation between regions as well as individuals' preferences of regions. A larger $\MobilityPred_{u,r}^{(T)}$ indicates that user $u$ will be more likely to visit region $r$ in the next time slot.

\subsection{Formulation and Analysis}

We design time slot-based strategy for the D2D replication for social content. Let $\Strategy^{(T)}$ denote the strategies users are going to take for content replication in the D2D social content delivery in time slot $T$. In $\Strategy^{(T)}$, each entry $K_{u,c}^{(T)}$ is a strategy variable: $K_{u,c}^{(T)} = 1$ indicates that user $u$ will \emph{carry/cache} content $c$ for D2D delivery in time slot $T$, and $K_{u,c}^{(T)} = 0$ otherwise.

Our objective is then to find an assignment for users, to best match the regional popularities of the social content. To capture how much a strategy matches the current social popularity of content in the edge regions, we define $Y_c^{(T)}$ as a replication \emph{gain} for social content $c$, under a given strategy $\Strategy^{(T)}$, as follows:
\begin{equation}
	Y_c^{(T)} = \sum_{u \in \Users} \beta_u K_{u,c}^{(T)} \sum_{r \in \Regions}  \MobilityPred_{u,r}^{(T)} \SocialPopularity_{c,r}^{(T)},
\end{equation}
where $\Users$ is the set of users who can perform social content replication, and $\beta_u$ is the upload capacity that user $u$ can contribute in the D2D delivery. $\beta_u$ is regarded as an altruism index for a user, and its value can be set according to the incentive mechanism in the system. For example, if the system is willing to pay users with credits if they contribute their upload, a large $\beta_u$ is expected.
The rationale of the D2D replication gain is that $Y_c^{(T)}$ reflects a \emph{matching level} of users performing the replication for the predicted regional social popularities of the content items. A larger $Y_c^{(T)}$ indicates that requests are more likely to be successfully served by users. 

The problem is then formulated as an optimization to find how contents generated and shared are assigned to potential D2D \emph{peering} users, so that the overall replication gain can be maximized. Our formulation is as below:
\begin{align}
\max_{\Strategy^{(T)}} &\, \sum_{c \in \setC^{(T)}} Y_c^{(T)}, \\
\mbox{s.t.} &\, \sum_{c \in \setC^{(T)}} K_{u,c}^{(T)} \le B_u, \forall u \in \Users, \label{equ:cache-capacity}\\
 & \, \sum_{u} \MobilityPred_{u,r}^{(T)} K_{u,c}^{(T)} \beta_u \le \SocialPopularity_{c,r}^{(T)}, \forall r \in \Regions, c \in \setC^{(T)}, \label{equ:popularity-constrain}\\
\mbox{vars.} &\, \Strategy^{(T)},\nonumber
\end{align}
where $\setC^{(T)}$ is the set of content items generated or shared by users, $B_u$ is the replication capacity of user $u$. Constrain (\ref{equ:cache-capacity}) requires that replication of a user will not exceed its cache capacity. Constrain (\ref{equ:popularity-constrain}) requires that replication for a particular social content item will not exceed its popularity estimation.

\subsection{Algorithm and Implementation}

The problem is in nature hard to solve and centralized, thus we design a heuristic local algorithm to make replication decision using local information.

\subsubsection{User Replication Algorithm}

Our solution for users to carry out the propagation- and mobility-aware D2D replication is as follows. When a user is in an edge region, it is assigned to replicate content that will propagate over users that are nearby. Notice that the user itself may not be actually sharing in the propagation it replicates content for. In our design, a user can replicate the social content either by directly receiving it from other users close to it, or receiving it from the content server. The user selects a subset from all the candidate social content items to cache for D2D delivery only according to its local information.

The details of the local replication strategy are illustrated in Algorithm \ref{alg:replication}. $\setW_u^{(T-1)}$ denotes the set of content items user $u$ has already replicated previously. $\setC_u^{(T)}$ is defined as the set of candidate content items to replicate, which contains contents that are estimated to propagation to regions where $u$ is predicted to move across in the near future time slot:
\begin{equation}
	\setC_u^{(T)} \equiv \{c | \SocialPopularity_{c,r}^{(T)} > 0, \MobilityPred_{u,r}^{(T)}>0 \}.
\end{equation}
The set $\setZ \equiv \setW_u^{(T-1)} \bigcup \setC_u^{(T)}$ will then be the candidate content items for user $u$ to replicate for the next time slot. Let $\setW_u^{(T)}$ denote the set of content items that user $u$ chooses to replicate. To determine which content items to carry for D2D delivery, a coordinate server calculates a local replication gain for content items in $\setZ$. $c.gain \equiv \sum_{r \in \Regions} \MobilityPred_{u,r}^{(T)} \SocialPopularity_{c,r}^{(T)}$ for user $u$ to carry content $c$ (line \ref{line:gain}). After that, $B_u$ content items will be randomly chosen from set $\setZ$, where a content item has the probability of $\frac{c.gain}{\sum_{c' \in \setZ} c'.gain}$ to be chosen (line \ref{line:prob}).

Users that have replicated content items from others will then send notification to the coordinate server, which can help users find potential edge peers to download social content items. Notice that traffic for such information communication at coordinator servers is much smaller than the traffic of content delivery.

\begin{algorithm}[t]
	\caption{D2D content replication.}\label{alg:replication}
	\begin{algorithmic}[1]
		\Procedure{Regional Content Replication}{}
			\State Let $\setZ \gets \setW_u^{(T-1)} \bigcup \setC_u^{(T)}$
			\ForAll{content $c \in \setZ$}
				\State $c.gain \gets \sum_{r \in \Regions} \MobilityPred_{u,r}^{(T)} \SocialPopularity_{c,r}^{(T)}$ \label{line:gain}
			\EndFor
			\State Let $\setW_u^{(T)}$ contain at most $B_u$ content items randomly selected from $\setZ$, where each content has the probability $\frac{c.gain}{\sum_{c' \in \setZ} c'.gain}$ to be chosen \label{line:prob}
			\State User $u$ cache content items in $\setW_u^{(T)}$ and report it to the coordinate server
		\EndProcedure
	\end{algorithmic}
\end{algorithm}

\subsubsection{On-demand Service}

Users replicating social content will serve others in an on-demand way: a requesting user who receives the \emph{meta information} of a content in the social propagation will first contact the coordinator server, to find which users in the same region have replicated the content. If a set of replicating users are found, the user will randomly select one to download the content from. If the content cannot be downloaded from local peers, {\em e.g.}, due to limitation of peers' upload capacities, the user will turn to the original content servers to download it, guaranteeing that the user can download the requested content within a given delay.

\subsubsection{Complexity Analysis}

In our design, each user individually determines which content it will replicate using the algorithm above, based on the social popularity indices provided by the centralized server, which periodically calculates $\SocialPopularity_{c,r}^{(T)}, \forall c \in\setC^{(T)}, r \in Regions$. Though $|\setC^{(T)}| |\Regions|$ can be too large for a single server to maintain the social popularity indices for all content items, the calculation itself can scaled in a horizontal manner: (1) We partition the content items into several subsets. (2) Content items in each subset will be handled by one coordinate server, which maintains the social popularity index. (3) Users in the edge networks will retrieve the social popularity indices from different servers according to the content.

\section{Performance Evaluation} \label{sec:evaluation}

In this section, we present our evaluation results of the propagation- and mobility-based replication strategy, based on the trace-driven experiments.

\subsection{Experiment Setup}

We implemented our algorithms in C++ in a simulator, based on an event-driven programming model, i.e., user mobility and social propagation activities, as well as network transmissions are simulated as events with action times and handlers \cite{montresor2009peersim}. The activities are driven by the real traces.

\textbf{Combining mobility and propagation behaviors.} We borrow the design idea from \cite{keranen2009one}, the ONE simulator for DTN, and map users in the social propagation traces to the users in mobility traces, i.e., the social behaviors and mobility behaviors from two traces are combined and assigned to one user in the simulation, following different correlation levels---note that we adjust the user mapping scheme in our experiments to vary the correlation between user mobility and social propagation and evaluate its impact on the effectiveness of our design.

\subsubsection{Mobility Behaviors}

In our experiments, two types of scenarios have been used to investigate user mobility patterns.

In our experiments, we also use an outdoor mobility dataset provided by Tencent Wi-Fi \cite{tencent}, a mobile app that asks users to submit how they use Wi-Fi networks. In particular, we have collected how $1.2$ million users move across the urban areas and associate with about $1$ million Wi-Fi hotspots in Shenzhen over $10$ days. Using these traces, we are able to infer user mobility in outdoor areas.

\begin{figure}[!t]
    \centering
        \includegraphics[width=0.9\linewidth]{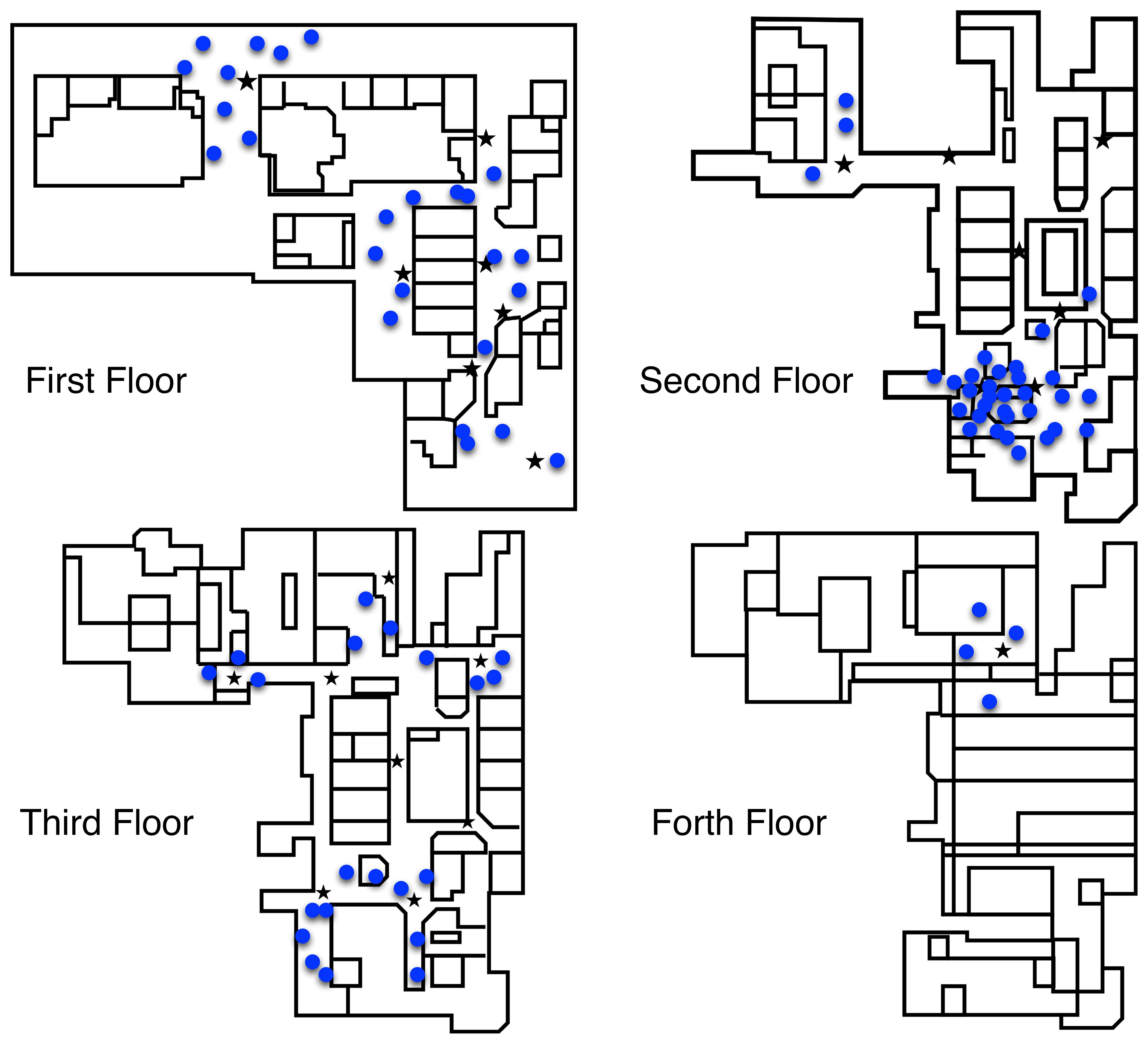}
    \caption{Illustration of locations of users in one time slot in our experiments: stars are the locations of Wi-Fi APs (regions), and circles represent users.}
    \label{fig:layout}
\end{figure}

The indoor area (i.e., the shopping mall in our previous measurement study) to simulate typical indoor usage. An illustration of these locations are selected from $4$ floors in the building. As illustrated in Fig.~\ref{fig:layout}, a star represents a Wi-Fi AP covering a certain edge-network region, and a circle represents a user, who is associated with the closest AP.

In both scenarios, we divide the area into 100x100m$^2$ regions in which users can communicate to each other. In our simulation, over $10$k users' mobility traces are used in different areas. Each region has a popularity for users to visit according to the statistics of users staying in these regions. In Table \ref{tab:exp-param}, we present the important parameters used in our experiments.

\begin{table}[!t]
    \caption{Experiment parameters used in both scenarios.}
    \label{tab:exp-param}
    \begin{center}
    \begin{tabular}{|p{.3\linewidth}|p{.3\linewidth}|p{.3\linewidth}|}
        \hline
            Parameter     & Indoor & Outdoor \\
        \hline
			Social connections & $40$ & $40$ \\
			$\PostLambda$ & $[0.001,0.02]$ & $[0.001,0.02]$ \\
			Re-share latency & $10$ hours & 10 hours \\
			Area size & $40\times100\times100$m$^2$ & $5\times5$km$^2$ \\
			Crowdedness & $[0,15]$ & $[0,5]$ \\
        \hline
    \end{tabular}
    \end{center}
\end{table}

\subsubsection{Social Behaviors}

We also select users from the social propagation traces and randomly map them to the users in the mobility traces. First, we build social connections between these users according to the distribution of social connections in the real world recorded in our traces. In our experiments, each user has about $40$ social connections on average.

Then, we simulate the social propagation. Each time slot is set to be $5$ minutes, which is about the average time a user stays in a region according to our measurement study. The number of content items for a user $u$ to post in each time slot in the online social network follows a Poisson distribution, which has a probability mass function $f(k;\PostLambda)=\frac{\PostLambda^k e^{-\PostLambda}}{k!}$. $\PostLambda$ follows the distribution of the average number of content items posted by users in each time slot in the social propagation traces, in the range of $[0.001,0.02]$ per time slot (about $[0.2,5]$ per day), which is summarized from our traces. After a content item is posted/shared by a user, it will be re-shared by others. The average latency of the re-shares is $10$ hours.

In the online social network, users can reach the content generated by others through the social connections. We observe that the number of content items propagating over different social connections is different. Based on our previous observation that the distribution of social propagation intensity of different social connections follows a power-law distribution \cite{zhi-tpds-dispersing}, we assign the re-share probability to each of the social connections following a power-law distribution learnt from the social connections between the sampled users.



\subsection{Baselines}

In our study, we compare our design with baselines as follows.

$\rhd$ \emph{Movement-based} replication. In this scheme, contents are replicated only based on the mobility patterns of users. Many previous studies lie in this category. We implement the movement-based strategy according to the spatial mobility analysis surveyed in \cite{karamshuk2011human}. Our implementation of the movement-based scheme is as follows: (1) We build a contact measurement index for each user, i.e., a large index is assigned to a user who has contacted more other users; (2) Based on this contact measurement, we prioritize users with their contact indices; (3) When there is a content item to replicate, a user with a larger contact index is more likely to be selected to carry the content item. The rationale is that such replication scheme greedily makes use of peers that have large chance to distribute the content to more users.

$\rhd$ \emph{Popularity-based} replication. In this scheme, content items are replicated to peers according to their popularities. In conventional peer-assisted content delivery, such popularity-based approach has been widely used \cite{huang2008challenges}. In our implementation of the popularity-based replication, the strategies are as follows: (1) A user ranks the popularity of content items according to their requests received; (2) When there is other users moving around, the user will choose a content item randomly from its local storage (a more popular content item has a larger chance to be chosen), and ask other users to replicate the content items; (3) A user when receiving multiple replication requests from others, will also determine which ones to replicate according to their popularity.

\subsection{Metrics}

We study the impact of running factors in real world, including intensity of social propagation, crowdedness of users in the regions (e.g., the number of users in each 100x100m$^2$ region), and the impact of popularity distribution of social contents. We use the following metrics to verify our design: (1) D2D delivery fraction, which is the fraction of traffic load that our D2D mechanism carries over all traffic served by both users and servers for the whole system; (2) The delivery load distribution of users who perform the D2D delivery.

\subsection{Experiment Results}

\subsubsection{Performance Improvement}

\begin{figure*}[t]
	\centering
	\subfloat[Outdoor scenario] {
        \includegraphics[width=.45\linewidth]{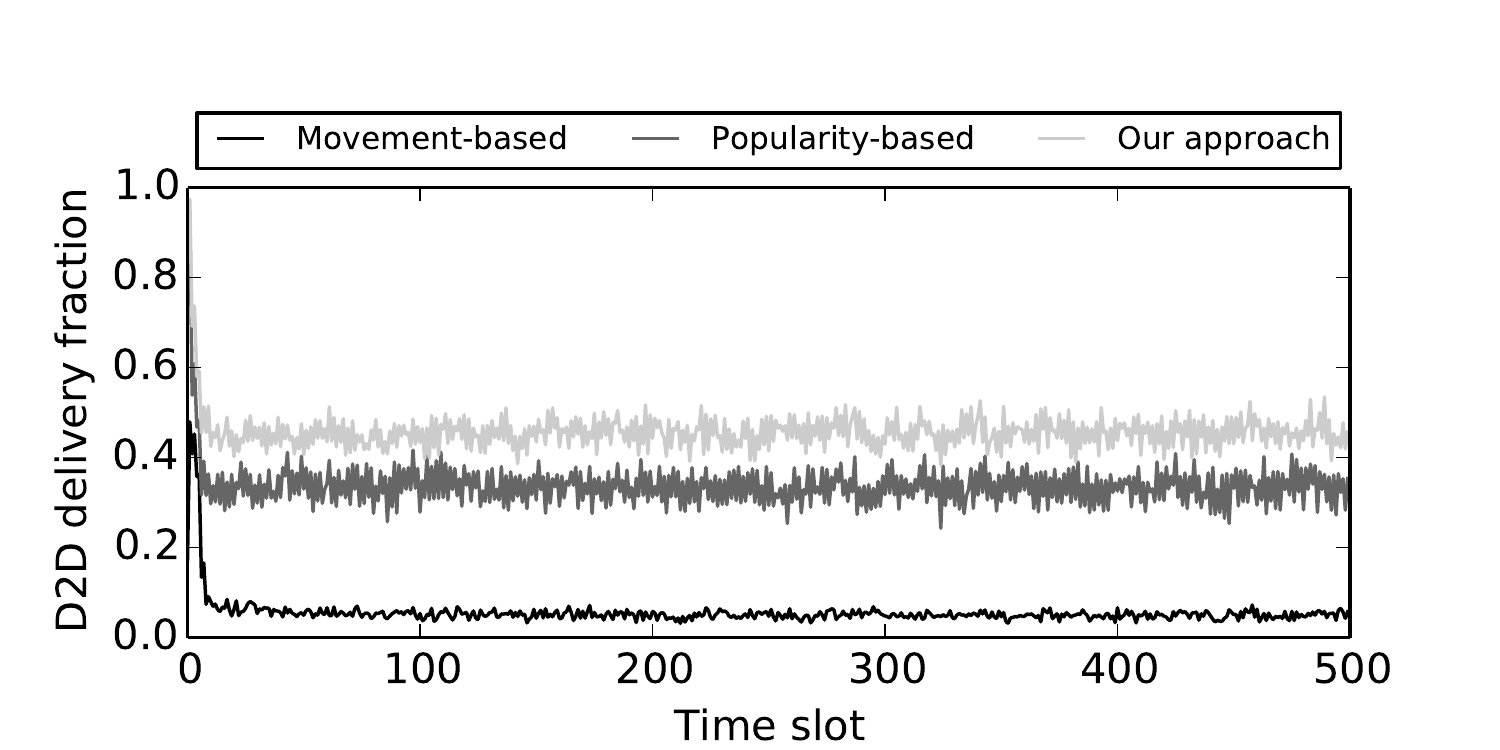}
        \label{fig:outdoor-cmp-over-time} }
    \subfloat[Indoor scenario] { 
        \includegraphics[width=.45\linewidth]{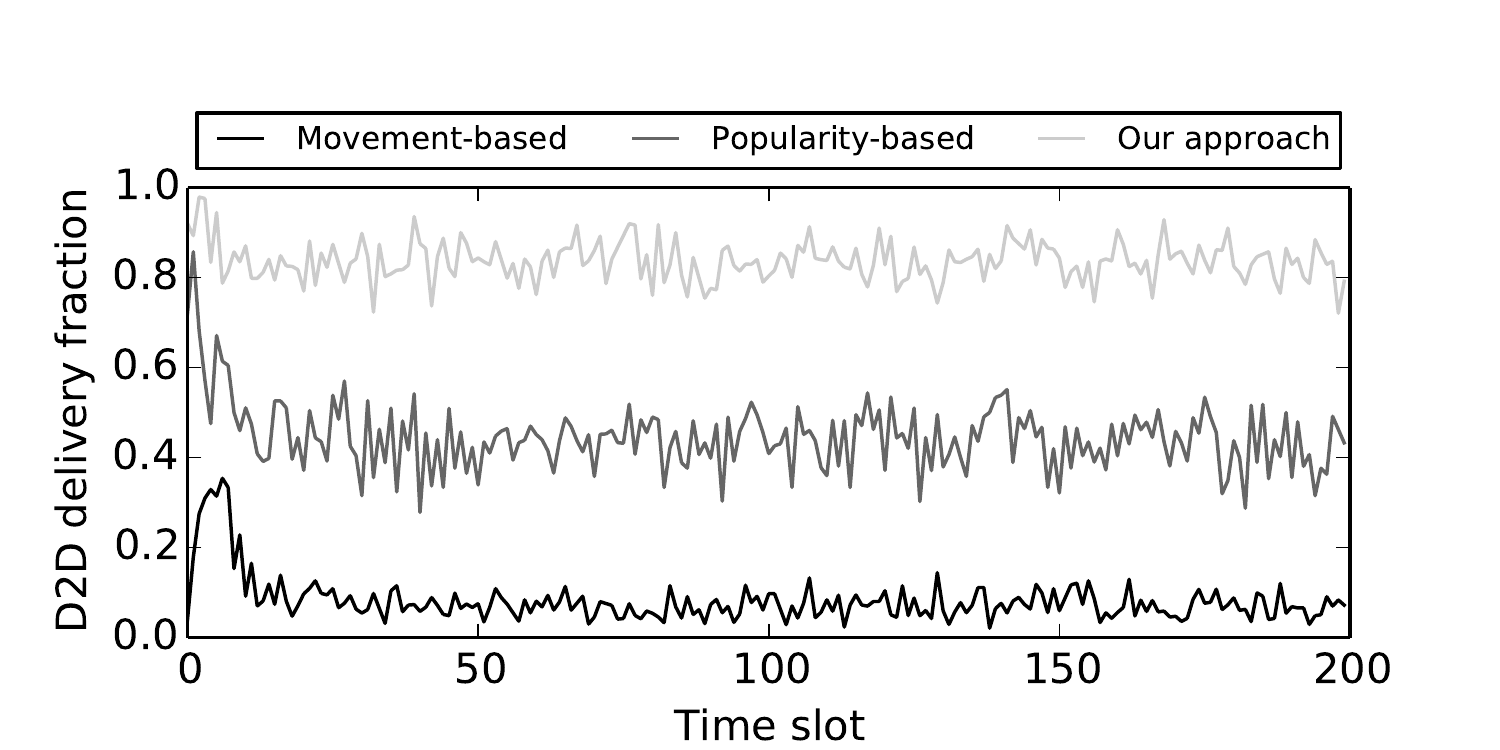}
        \label{fig:indoor-cmp-over-time}}
	\caption{Comparison of fraction of requests successfully delivered by D2D replication over time in the indoor scenario.}
	\label{fig:D2D-frac-cmp-overtime}
\end{figure*}

We study the effectiveness of our design, in terms of the amount of social contents that can be delivered by D2D replication, compared to the pure movement-based and pure popularity-based approaches. As illustrated in Fig.~\ref{fig:D2D-frac-cmp-overtime}, the curves are the fraction of D2D delivered content over time, in both outdoor and indoor scenarios. 

Our results are as follows. (1) In the outdoor scenario, the D2D delivery performance of our design is slightly better than the popularity-based approach, and about $45\%$ of content items are delivered by D2D delivery by both strategies, as illustrated in Fig.~\ref{fig:outdoor-cmp-over-time}. (2) In the indoor scenario as illustrated in Fig.~\ref{fig:indoor-cmp-over-time}, our design can improve by $4$ times against the pure movement-based approach, and $2$ times against pure popularity-based approach. This result indicates that in edge-network content delivery, our proposal based on regional propagation and mobility predictions is more capable to make full utilization of edge-network users' peering resources, especially when propagation is intensive in the local area.

\subsubsection{Users' D2D Contribution}

\begin{figure}[!t]
		\centering
			\includegraphics[width=.9\linewidth]{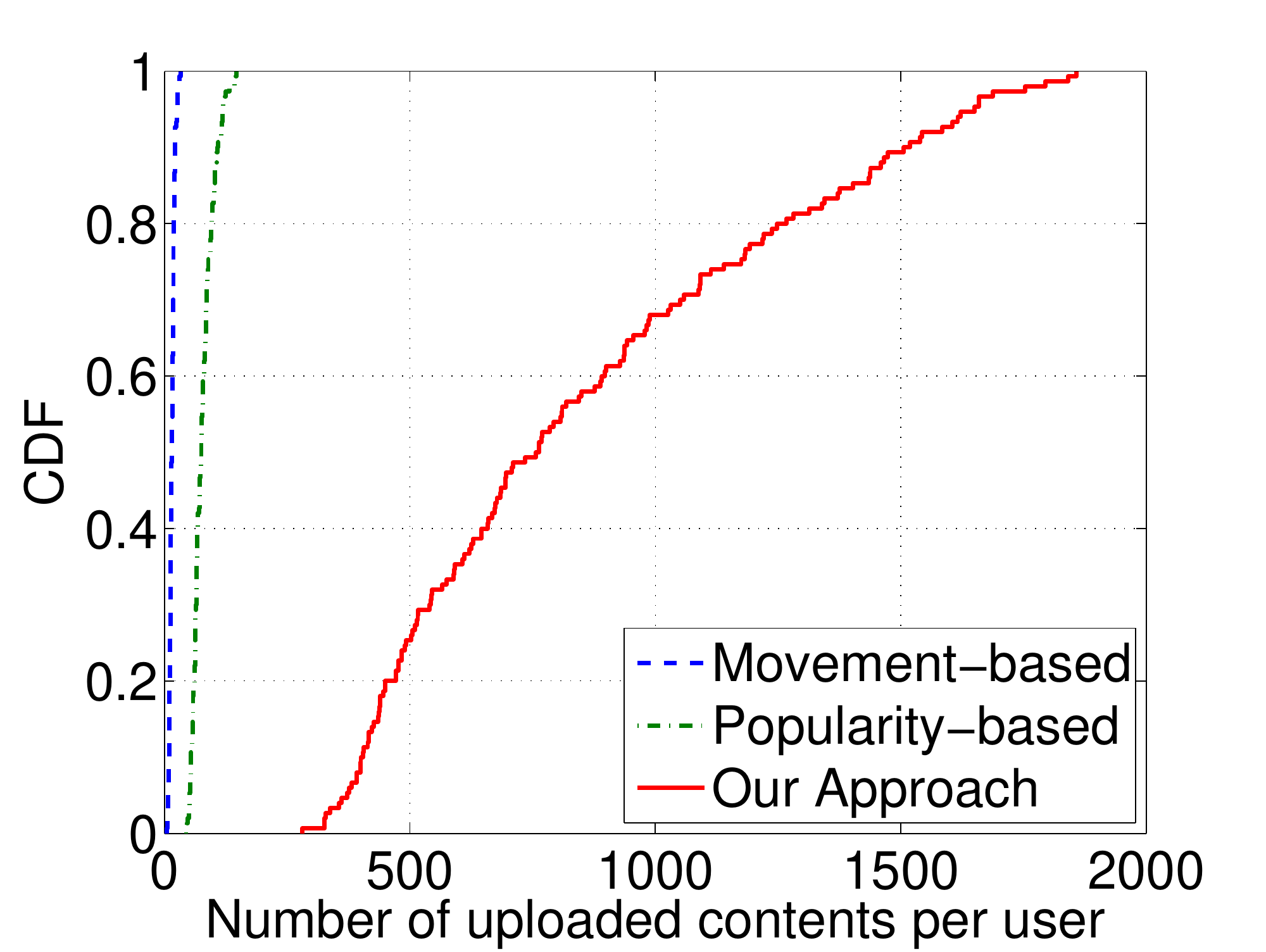}
		\caption{CDF of users' contribution in D2D social content delivery in the indoor scenario.}
		\label{fig:peer-contribution-cdf}
\end{figure}

We investigate the actual contributions of users, in terms of the number of social contents served by them. In Fig.~\ref{fig:peer-contribution-cdf}, each curve is the CDF of users' contribution (the number of uploaded social contents). We observe that in our design, peers serve much more social contents to their neighbors than the other two schemes. In particular, in our design, the users who can contribute the least are more utilized than other strategies. In our experiments, a fraction of larger than $50\%$ of the content items are downloaded by users that are not requested by themselves, but only for other users. This indicates that good incentive mechanisms are required in such D2D systems.

\subsubsection{Social Propagation Intensity}

\begin{figure}[t]
	\centering
		\includegraphics[width=.9\linewidth]{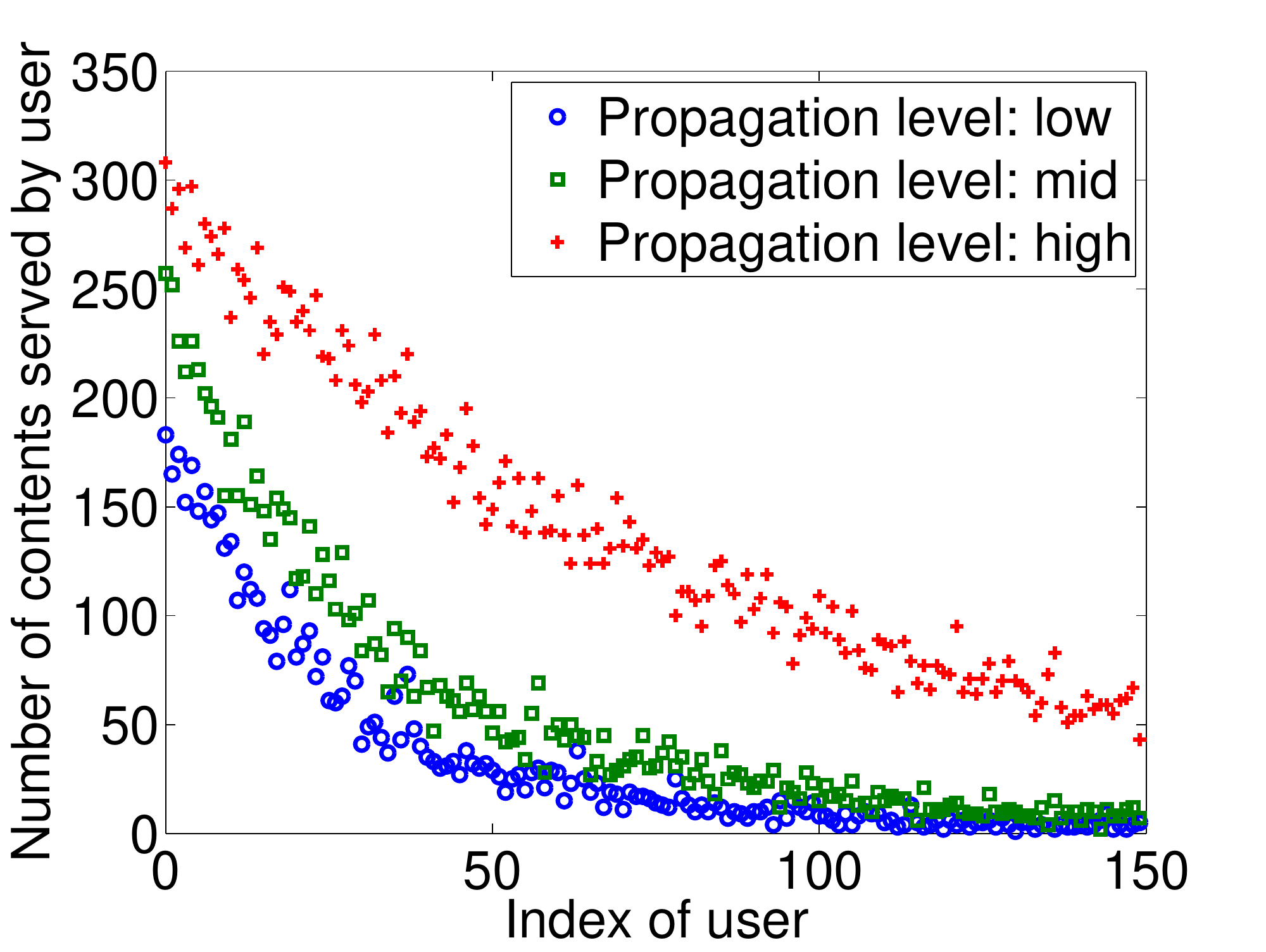}
	\caption{Users' D2D contribution under different social propagation intensity in our design.}
	\label{fig:propagation-impact}
\end{figure}

We study the impact of social propagation intensity by varying the number of social contents generated and shared by users per time slot. As illustrated in Fig.~\ref{fig:propagation-impact}, each sample is the contribution of a particular user versus her user index. In the high propagation level, the number of social contents generated or shared by users per time slot is around $10$; in the mid propagation level, the number of that is around $6$; and in the low propagation level, the number is around $2$. We observe that our design can well adapt to the social propagation intensity, all under the same storage and upload capacity. We also observe that when the level of propagation intensity is high, users tend to have more similar contribution.

\subsubsection{Regional Crowdedness}

\begin{figure}[t]
	\centering
		\includegraphics[width=0.9\linewidth]{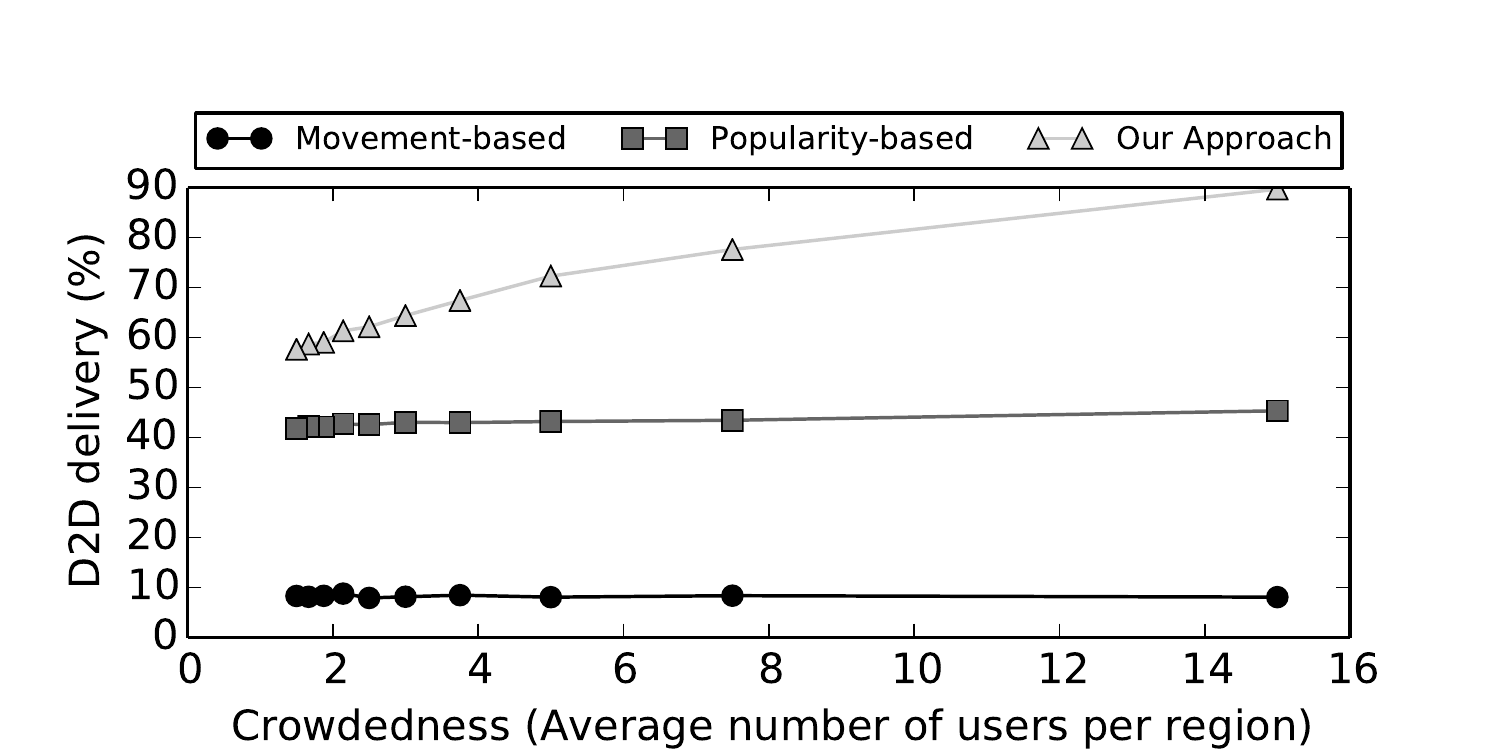}
	\caption{Comparison of fraction of requests successfully delivered by D2D replication versus crowdedness.}
	\label{fig:D2D-frac-cmp-vs-crowdedness}
\end{figure}

D2D content delivery is in nature based on the device-to-device communication; we thus evaluate the impact of \emph{crowdedness} of regions, which captures the average number of users per region. By varying the number of users in our experiments, we are able to change the regional crowdedness. As illustrated in Fig.~\ref{fig:D2D-frac-cmp-vs-crowdedness}, the curves represent the D2D delivery fraction against the average number of users per region during an experiment. We observe that different from the movement-based and popularity-based approaches, performance of our design is actually sensitive to the change of crowdedness. In particular, the D2D delivery fraction in our design increases when the number of users per region increases, especially in regions where there are few users.

\subsubsection{Distance between Friends}

By mapping social propagation users to mobility users in different scenarios, we have different distances between friends. We calculate the average distance between each pair of friends and study the impact of the average distance on the performance of these algorithms. As illustrated in Table \ref{tab:impact-distance}, we observe that such D2D delivery performs well when social propagation happens between friends that are close to each other, which is consistent with our previous results. When friends are moving in a city level, i.e., the average distance is above $5$km, the D2D delivery drops to about $30\%$, which is only slightly better than the popularity-based approach.

\begin{table}
    \caption{Impact of average distance among friends on D2D delivery fraction.}
    \label{tab:impact-distance}    
    \begin{center}
    \begin{tabular}{|c|c|}
        \hline \textbf{Average distance (m)} & \textbf{D2D delivery fraction} \\
        \hline $[0,500)$  & $0.76$ \\
        \hline $[500,1500)$  & $0.65$ \\
        \hline $[1500,2500)$  & $0.59$ \\
        \hline $[2500,5000)$  & $0.48$ \\
        \hline $[5000,\inf)$  & $0.31$ \\
        \hline
    \end{tabular}
    \end{center}
\end{table}

\subsubsection{Correlation between Social Behavior and Mobility Behavior}

In real world, the social propagation including content generation and propagation behaviors are not independent on the user mobility, e.g., users may post more content items when they are waiting for bus. We investigate the impact of the correlation between social behaviors and mobility behaviors, by using different user mapping schemes as follows. (1) Independent: we randomly map social users (users collected from the social network traces) to mobility users (users collected from the mobility traces); (2) Social rank: We rank users according to their social behavior intensity (i.e., the number of content posts and re-shares), and map the ranked users to mobility users in a random rank of regions; (3) Social-mobility rank: We rank users in the social network according to the propagation intensity, and rank users from the mobility traces according their mobility intensity; then we map the ranked social users to the ranked mobility users.

\begin{figure}[!t]
    \centering
        \includegraphics[width=.9\linewidth]{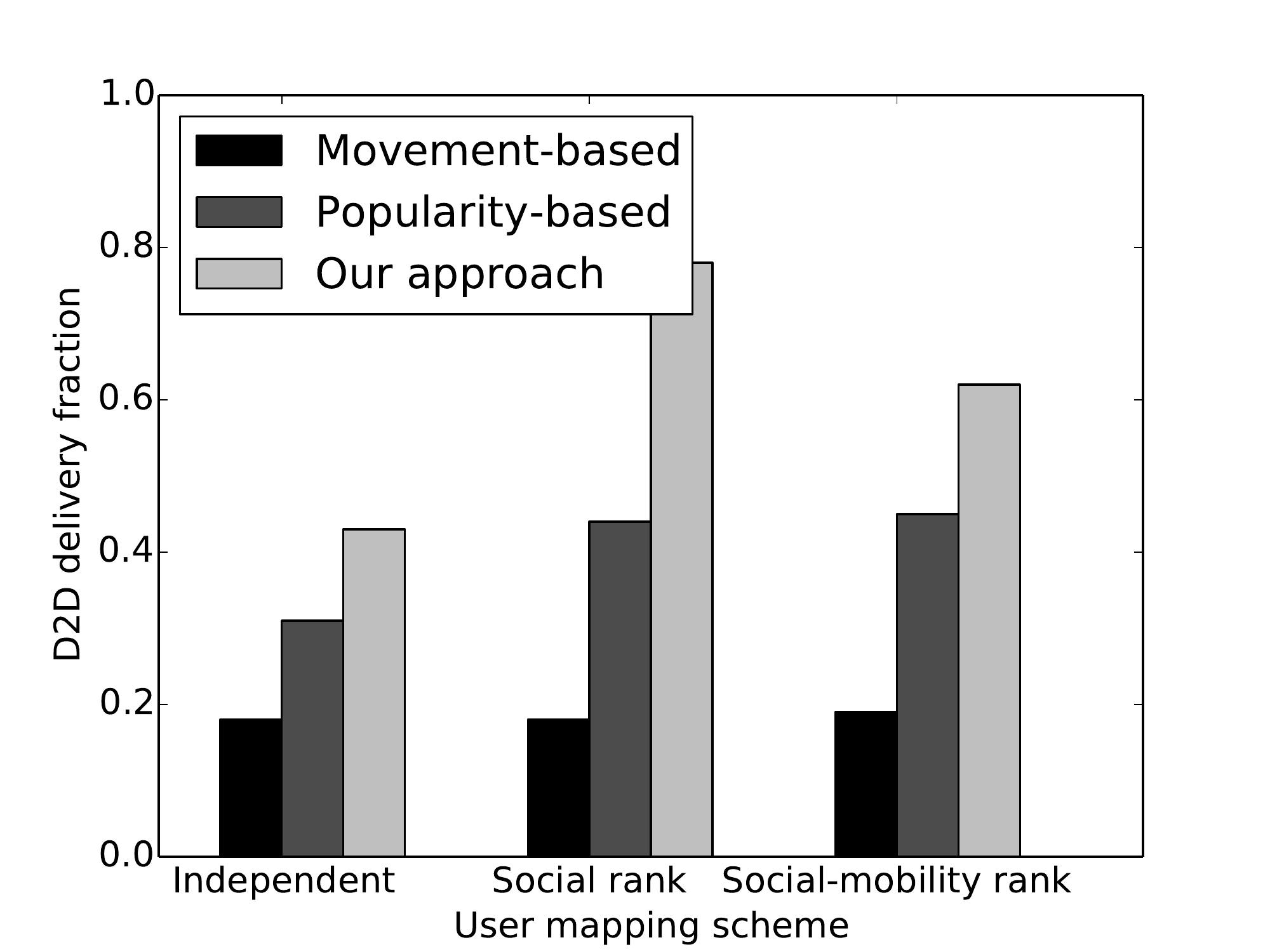}
    \caption{Impact of social behavior and mobility behavior.}
    \label{fig:corr-impact}
\end{figure}

As illustrated in Fig.~\ref{fig:corr-impact}, we plot the D2D delivery fraction of different strategies under different user mapping schemes. We observe that our approach is more sensitive to different user mapping schemes than the movement-based and popularity-based approaches. We also observe that our design performs well when social behaviors happen in nearby regions, and social propagation and mobility are not highly correlated. These observations indicate that the performance of our design is affected by whether the social behaviors are independent on mobility behaviors. In particular, the performance is better when social behaviors are not uniformly distributed in different regions.

\subsubsection{Content Popularity}

\begin{figure}[t]
	\centering
		\includegraphics[width=.90\linewidth]{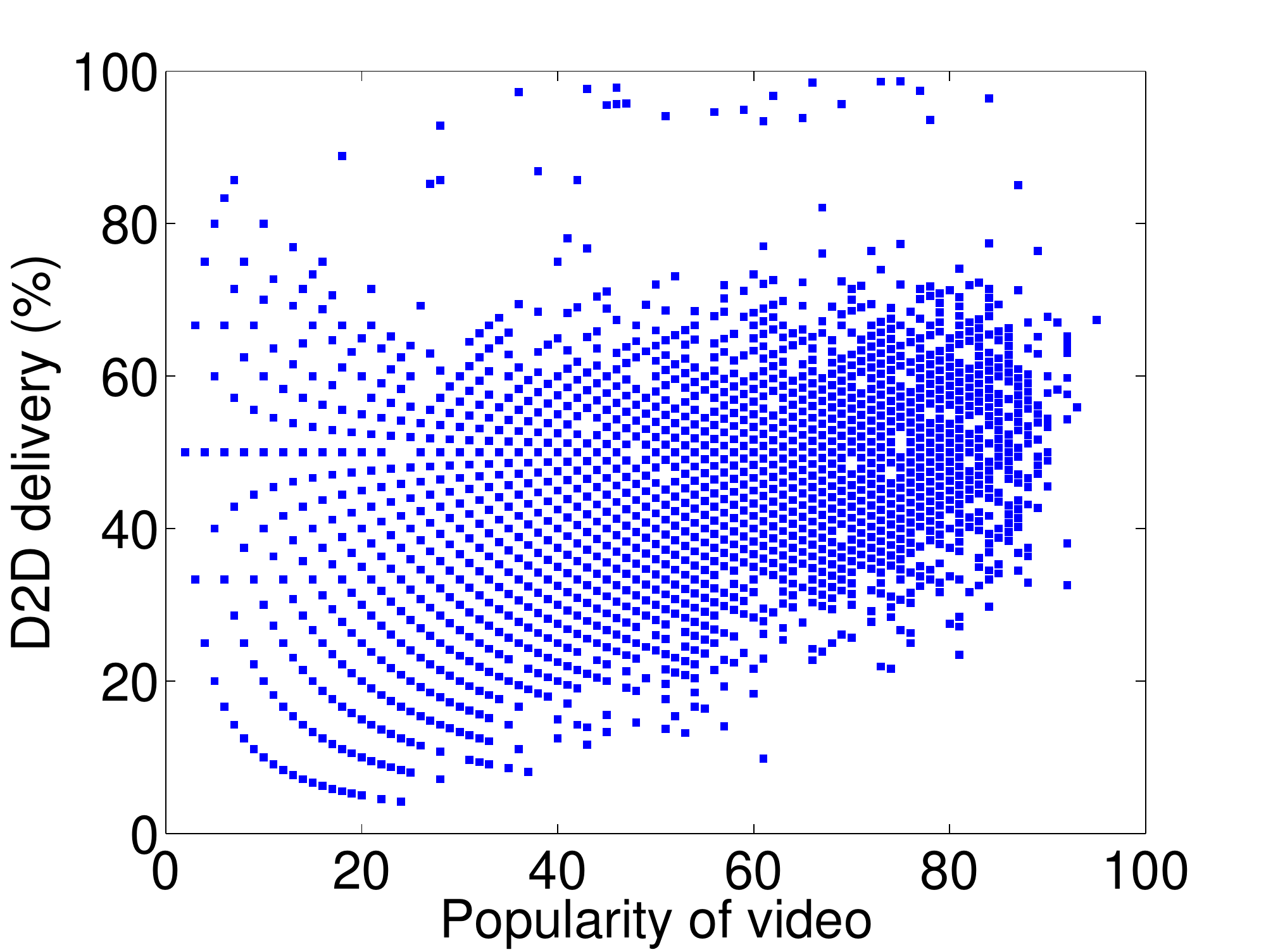}
	\caption{Fraction of D2D delivery versus the popularity of contents.}
	\label{fig:peerfrac-vs-popularity}
\end{figure}

We study the impact of the popularity of social contents, which is calculated as the number of all previous requests from users for a content. In Fig.~\ref{fig:peerfrac-vs-popularity}, each sample is the D2D delivery fraction of a particular content during the experiments versus the popularity of the content. In general, we observe that our design can handle social contents with different popularities. In particular, we observe a slight trend that the D2D delivery fraction is larger when a content's popularity is higher. The reason is that, it is more possible for contents of high popularity (which are likely to be already cached by many users) to be better replicated.

\subsubsection{Fraction of Content Handled}

\begin{figure}[!t]
    \centering
        \includegraphics[width=.9\linewidth]{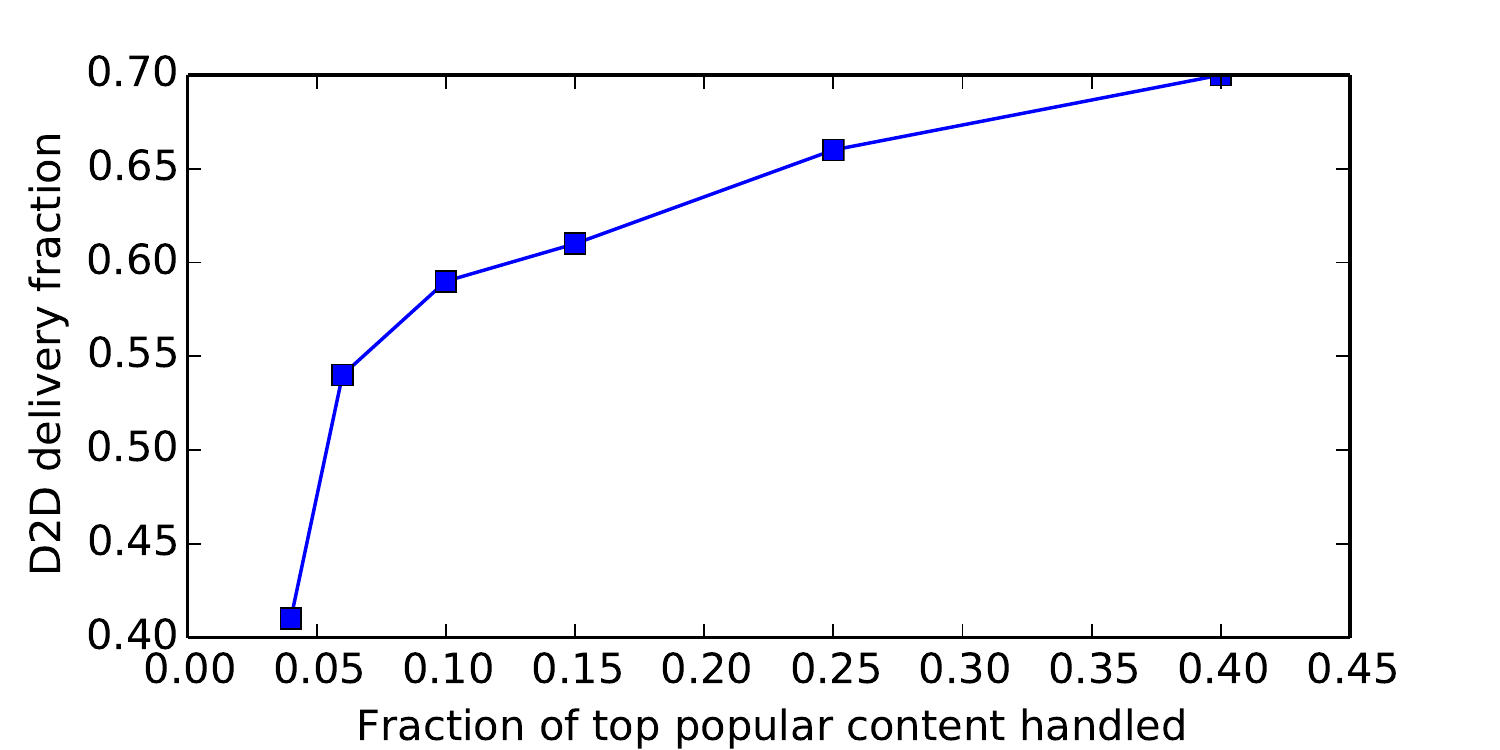}
    \caption{Impact of fraction of the top popular content handled.}
    \label{fig:impact-of-top-content}
\end{figure}

In a practical implementation, the system might not be able to monitor information about all the content items shared between users. We thus investigate the impact of the fraction of content items handled by the D2D delivery. We vary the fraction of the most popular content items that are considered to be delivered by peers, and let the server serve the other content items. In Fig.~\ref{fig:impact-of-top-content}, we plot the D2D delivery fraction versus the fraction of top popular content handled by D2D. We observe that due to the high skewness of the popularity distribution of social content, our design can achieve relatively high performance by only handle a small fraction of the most popular content.

\section{Concluding Remarks} \label{sec:conclusion}

The massive number of user-generated bandwidth-intensive social contents and their in nature highly-skewed popularity distribution, make conventional content delivery based on a static and hierarchical infrastructure inefficient. In particular, it is hard to serve users with network resources right close to them for every single social content shared. Motivated by the development of device-to-device communication, we propose a D2D replication for social contents. Based on extensive large-scale measurement studies, we find the local sharing and delay-tolerant characteristics of social content sharing, and the regional propagation and mobility patterns of users. Based on those insights, we design regional propagation and mobility predictive models, to estimate \emph{where a social content may propagate to during social propagation} and \emph{which user can replicate it on her move}. We formulate the problem and design a heuristic algorithm based on only historical and local information to solve it. Trace-driven experiments further verify the effectiveness of our design, which not only outperforms conventional movement-based and popularity-based approaches by $2-4$ times, but also is adaptive to different levels of social propagation intensity, regional crowdedness and content popularity.

\bibliographystyle{plain}
\bibliography{mylib}

\end{document}